\documentclass[useAMS,usenatbib,usegraphicx]{mn2e}
\usepackage{amsmath,amssymb}

\title[A wide-field survey of globular clusters - I]{A {\it
Subaru/Suprime-Cam} wide-field survey of globular cluster populations
around M87 - I: Observation, data analysis, and luminosity function}

\author[N. Tamura et al.]{Naoyuki
Tamura$^{1}$\thanks{E-mail:naoyuki@naoj.org}\thanks{Curent address:
Subaru Telescope, National Astronomical Observatory of Japan, 650 North
A'ohoku Place, Hilo, HI 96720, USA}, Ray M. Sharples$^{1}$, Nobuo
Arimoto$^{2}$, Masato Onodera$^{2, 3}$, \newauthor Kouji Ohta$^{4}$, \&
Yoshihiko Yamada$^{2}$ \\
$^{1}$Department of Physics, University of Durham, South Road, Durham,
DH1 3LE, United Kingdom \\
$^{2}$National Astronomical Observatory of Japan, Mitaka, Tokyo
181-8588, Japan \\
$^{3}$Department of Astronomy, School of Science, University of Tokyo,
Tokyo 113-0033, Japan \\
$^{4}$Department of Astronomy, Faculty of Science, Kyoto University,
Kyoto 606-8502, Japan
}
 
\begin{document}

\date{}
 
\pagerange{\pageref{firstpage}--\pageref{lastpage}} \pubyear{2006}
 
\maketitle
 
\label{firstpage}

\begin{abstract}

In this paper and a companion paper, we report on a wide-field imaging
survey of the globular cluster (GC) populations around M87 carried out
with Suprime-Cam on the 8.2m Subaru telescope. Here we describe the
observations, data reduction, and data analysis and present luminosity
functions of GC populations around M87 and NGC 4552, another luminous
Virgo elliptical in our survey field. The imaging data were taken in the
$B$, $V$, and $I$ bands with a sky coverage of $2^{\circ} \times
0_{\cdot}^{\circ}5$ extending from the M87 centre out to $\sim$ 0.5
Mpc. GC candidates were selected by applying a colour criterion on the
$B-V$ and $V-I$ diagram to unresolved objects, which greatly reduces 
contamination. The data from control fields taken with
Subaru/Suprime-Cam were also analyzed for subtraction of contamination
in the GC sample. These control field data are compatible with those in
the M87 field in terms of the filter set ($BVI$), limiting magnitudes,
and image quality, which minimizes the possibility of introducing any
systematic errors into the subtractive correction.
We investigate GC luminosity functions (GCLFs) at distances $\leq
10^{\prime}$ ($\lesssim$ 45 kpc) from the host galaxy centre in
detail. By fitting Gaussians to the GCLFs, the $V$-band turnover
magnitude ($V_{\rm TO}$) is estimated to be $23.62 \pm 0.06$ mag and
$23.56 \pm 0.20$ mag for the GC population in M87 and NGC 4552,
respectively. The GCLF is found to be a function of GC colour; $V_{\rm
TO}$ of the red GC subpopulation ($V-I > 1.1$) is
fainter than that of the blue GC subpopulation ($V-I \leq 1.1$) in both
M87 and NGC 4552, as expected if the colour differences are primarily
due to a metallicity effect and the mass functions of the two
subpopulations are similar. The radial dependence of the GCLF is also
investigated for the GC population in M87. The GCLF of each
subpopulation at $1^{\prime} \leq R \leq 5^{\prime}$ is compared with
that at $5^{\prime} \leq R \leq 10^{\prime}$ but no significant trend
with distance is found in the shape of the GCLF.
We also estimate GC specific frequencies ($S_N$) for M87 and NGC 4552.
The $S_N$ of the M87 GC population is estimated to be 12.5 $\pm$ 0.8
within 25$^{\prime}$.
The $S_N$ value of the NGC 4552 GC population is estimated to be 5.0
$\pm$ 0.6 within $10^{\prime}$.

\end{abstract}


\section{INTRODUCTION}\label{intro}

Globular clusters (GCs) are homogeneous stellar systems containing stars
with a single age and metallicity, which are in principle simpler to
interpret than photometric and spectroscopic observations of the
integrated stellar light of a galaxy. GCs are therefore considered to be
powerful probes with which to understand the star formation and chemical
enrichment history of their host galaxy.
One of the basic findings from observations about GC populations in
luminous galaxies is that while thousands of GCs are associated with
luminous elliptical galaxies, a significantly smaller number of GCs
exist around spiral galaxies with similar luminosities (e.g., Harris
1991; Barmby 2003). This indicates that the specific frequency of GCs
($S_N$), which is considered to be related to the relative efficiency of
GC formation and/or survival compared to galactic halo/bulge stars,
depends on galaxy morphology. In fact, $S_N$ has also been suggested to
be correlated with local galaxy density, with galaxies in denser
environments having larger $S_N$ values (West 1993). The fact that this
trend appears to exist even when the sample of galaxies is restricted to
ellipticals suggests that GC formation efficiency is more physically
linked with galaxy environment. One possibility to explain this
observation is biased GC formation in galaxies inhabiting denser
environments (West 1993; Blakeslee 1999; McLaughlin
1999). Alternatively, a substantial number of GCs in a luminous galaxy
may have an external origin; GCs could be captured from other galaxies
through galaxy interactions and accrete onto luminous galaxies in
clusters, which would then enhance $S_N$ values as observed.

Recent studies of GCs in the central regions of luminous ellipticals
conducted with the {\it Hubble Space Telescope} (HST) have revealed that
many luminous ellipticals have bimodal or multimodal colour
distributions of GCs (Gebhardt \& Kissler-Patig 1999; Larsen et al.
2001; Brodie et al. 2005). It is found that the mean colours of both red
(metal-rich) and blue (metal-poor) GC subpopulations are correlated with
the host galaxy luminosities and colours (Larsen et al. 2001; Strader,
Brodie \& Forbes 2004; Strader et al. 2005; Peng et al. 2006), which
will be important constraints on the proposed scenarios for the
formation and evolution of GC population such as multiphase collapse
scenario (Forbes, Brodie \& Grillmair 1997), merger scenario (Ashman \&
Zepf 1992), hiearchical merging scenario (Beasley et al.  2002), and
accretion scenario (C\^{o}t\'{e}, Marzke \& West 1998). The small field
of view of the HST, however, does not allow one to collect GC
populations in the outer halo of a galaxy and investigate their spatial
structures, which will also be key pieces of the puzzle (e.g., Moore et
al. 2005). Much wider-field studies (e.g., out to $\sim$ 100 kpc from
the galaxy centre) of GC populations therefore need to be performed
using ground-based data (Rhode \& Zepf 2001, 2004; Dirsch et al. 2003;
Bassino et al.  2006). Studying GC populations at large distances from
the host galaxy is of great importance because the outer halo of the
host galaxy, and even the intergalactic space, are presumed to be large
reservoirs of blue and metal-poor GCs in the accretion scenario.

\setlength{\tabcolsep}{1.5mm}
\begin{table*}
\centering
\begin{minipage}{150mm}
\begin{tabular}{lccccccccc} \hline\hline
\multicolumn{1}{c}{Field ID}     & 
\multicolumn{3}{c}{Field centre} &
\multicolumn{3}{c}{Integration time ($m_{\rm lim}$)} & 
\multicolumn{3}{c}{Seeing size} \\[2pt]
&
&
&
&
\multicolumn{3}{c}{[sec (mag)]}  & 
\multicolumn{3}{c}{[arcsec]} \\
\multicolumn{1}{c}{(1)} & 
\multicolumn{3}{c}{(2)} &
\multicolumn{3}{c}{(3)} & 
\multicolumn{3}{c}{(4)} \\ \hline
& 
$\alpha$(J2000) &
$\delta$(J2000) &
$b$(J2000)      &
$B$ & 
$V$ & 
$I$ & 
$B$ & 
$V$ & 
$I$ \\
   
Field 1       & $12^{h}31^{m}18^{s}_{\cdot}4$ & $12^{\circ}29^{\prime}13^{\prime\prime}$ &
74$^{\circ}_{\cdot}$6 &
3680 (25.6) & 1350 (25.1) & 3480 (24.6) & 1.8 & 1.0 & 1.0 \\
Field 2       & $12^{h}33^{m}42^{s}_{\cdot}8$ & $12^{\circ}27^{\prime}37^{\prime\prime}$ &
74$^{\circ}_{\cdot}$8 &
2640 (25.9) & 1350 (25.2) & 3690 (24.7) & 1.2 & 1.0 & 1.0 \\
Field 3       & $12^{h}36^{m}08^{s}_{\cdot}8$ & $12^{\circ}24^{\prime}35^{\prime\prime}$ &
74$^{\circ}_{\cdot}$9 &
1800 (25.7) & 1350 (25.2) & 2640 (24.5) & 1.5 & 1.0 & 1.1 \\
Field 4       & $12^{h}38^{m}33^{s}_{\cdot}2$ & $12^{\circ}24^{\prime}35^{\prime\prime}$ &
75$^{\circ}_{\cdot}$0 &
2280 (26.1) & 1350 (25.2) & 4380 (24.8) & 1.1 & 1.0 & 1.0 \\ \hline
HDF-N         & $12^{h}36^{m}46^{s}_{\cdot}7$ & $62^{\circ}11^{\prime}50^{\prime\prime}$ &
54$^{\circ}_{\cdot}$8 &
6000 (26.6) & 4800 (25.5) & 4200 (25.2) & 0.8 & 1.1 & 0.9 \\
Lockman Hole  & $10^{h}35^{m}55^{s}_{\cdot}2$ & $57^{\circ}42^{\prime}18^{\prime\prime}$ &
51$^{\circ}_{\cdot}$3 &
6000 (26.9) & 4800 (26.0) & 3600 (24.8) & 1.0 & 1.1 & 1.3 \\ \hline
\end{tabular}

\caption{Observation log. Data for HDF-N and Lockman Hole are retrieved
from SMOKA. Col. (3): The integration times were calculated including
data taken under non-photometric conditions on the first night (hence
the interpretation is not straightforward). Number in the parentheses
indicates 50 \% completeness to point sources estimated with artificial
star test. Col. (4): The seeing sizes are estimated in the stacked
images.} \label{basic}

\end{minipage}
\end{table*}

\begin{figure*}
 \begin{center}
  \includegraphics[height=11cm,keepaspectratio,clip]{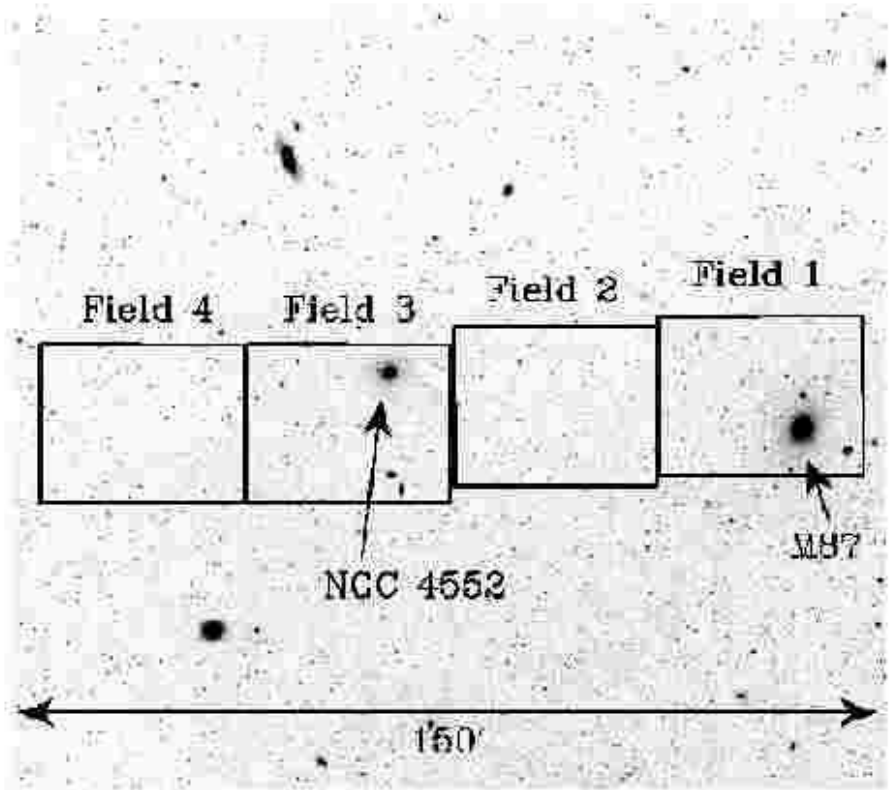}
  \caption{Locations and IDs of the observed fields. Each box indicates
  one SCam field of view. The background is a DSS image. North is up and
  East is left.}  \label{deff}
 \end{center}
\end{figure*}

In this paper and a companion paper (Tamura et al. 2006; Paper II
hereafter), we report a wide-field imaging survey of the GC populations
around M87 conducted with Suprime-Cam on the Subaru telescope. Several
moderately wide-field studies of GCs surrounding M87 have already been
carried out using photometry (Strom et al. 1981; McLaughlin, Harris, \&
Harris 1994; Harris, Harris, \& McLaughrin 1998; Hanes et al. 2001) and
spectroscopy (Cohen \& Rizhov 1997; Cohen, Blakeslee \& Rizhov 1998;
Kissler-Patig \& Gebhardt 1998; C\^{o}t\'{e} et al. 2001), but all of
these studies explored only the regions $\lesssim 10^{\prime}$
($\lesssim$ 45 kpc or 5 $r_e$) from M87. In contrast, the area of our
survey is approximately $2^{\circ} \times 0_{\cdot}^{\circ}5$ (560 kpc
$\times$ 140 kpc) extending from the centre of M87 out to $\sim$ 0.5
Mpc, which is the widest survey yet undertaken of the GC populations in
luminous galaxies.
In this paper, we focus on describing the observations, data reduction
(\S~\ref{obsredcal}) and data analyses such as selection and photometry
of GC candidates, incompleteness correction, and subtraction of
foreground and background contamination in the GC candidates
(\S~\ref{analyses}). We derive GC luminosity functions around M87 and
NGC 4552, another Virgo luminous elliptical galaxy in our survey field,
and estimate the global GC specific frequencies of these luminous
ellipticals in \S~\ref{resultsanddiscussions}. We investigate colour
distributions and spatial distributions of GC candidates in Paper II.
We adopt distances of 16.1 Mpc (distance modulus of 31.03) to M87, and
15.4 Mpc (distance modulus of 30.93) to NGC 4552, based on measurements
using the surface brightness fluctuation method (Tonry et al. 2001). An
angular scale of 1$^{\prime}$ corresponds to 4.7 kpc and 4.5 kpc at the
distance of M87 and NGC 4552, respectively.

\section{OBSERVATIONS, DATA REDUCTIONS, AND CALIBRATIONS}\label{obsredcal}

\subsection{The M87 Fields}\label{m87fields}

\subsubsection{Observations and data reductions}\label{m87obsred}

Imaging observations were performed on 17 and 18 March 2004, with
Suprime-Cam (SCam) (Miyazaki et al. 2002) on the Subaru telescope. SCam
is a mosaic CCD camera with 10 2K$\times$4K CCD chips and the field of
view is approximately $34^{\prime} \times 27^{\prime}$ on the sky. The
pixel scale is $0_{\cdot}^{\prime\prime}2$ pixel$^{-1}$. In this
observing program, a field of approximately 1 square degree
($136^{\prime} \times 27^{\prime}$) extending from M87 towards the east
was covered by 4 telescope pointings through $B$-, $V$-, and $I$-band
filters. The field IDs and locations are shown in Fig. \ref{deff} and
the observation log is presented in Table \ref{basic}. Each field was
observed with the telescope dithered by $\sim 5^{\prime\prime}$. Since
this dithering scale is smaller than the gap between the CCDs, the 10
CCD frames are not mosaiced into one continuous frame but are reduced
and analyzed individually. A typical exposure time of one frame is 360
sec, 270 sec, and 240 sec in $B$, $V$, and $I$ band, respectively;
several frames were co-added to give the total exposure times listed in
Table \ref{basic}.  On the first night, the sky condition was
non-photometric and the transparency was highly variable. On the second
night, it was much better but was still hazy with a little variation.
We therefore scale the data taken on the first night by shifting the
magnitude zeropoints to match with those of the data on the second
night, calibrate the reduced data based on the standard stars taken on
the second night, and check the calibration using GC photometry in the
literature (see next section for details). Typical seeing sizes during
the observations were $1_{\cdot}^{\prime\prime}5$ in $B$ band and $\sim
1_{\cdot}^{\prime\prime}0$ in $V$ and $I$ bands.

\begin{figure}
 \begin{center}
  \includegraphics[width=7.5cm,keepaspectratio]{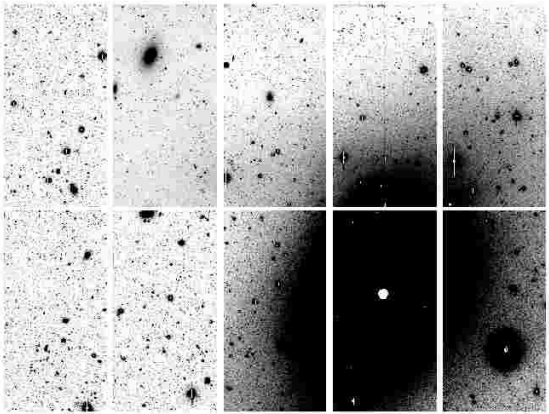}
 \end{center}
  \caption{Reduced data for Field 1 ($V$ band). North is up and East is
  left.}  \label{f1}
\end{figure}

\begin{figure}
 \begin{center}
  \includegraphics[width=7.5cm,keepaspectratio]{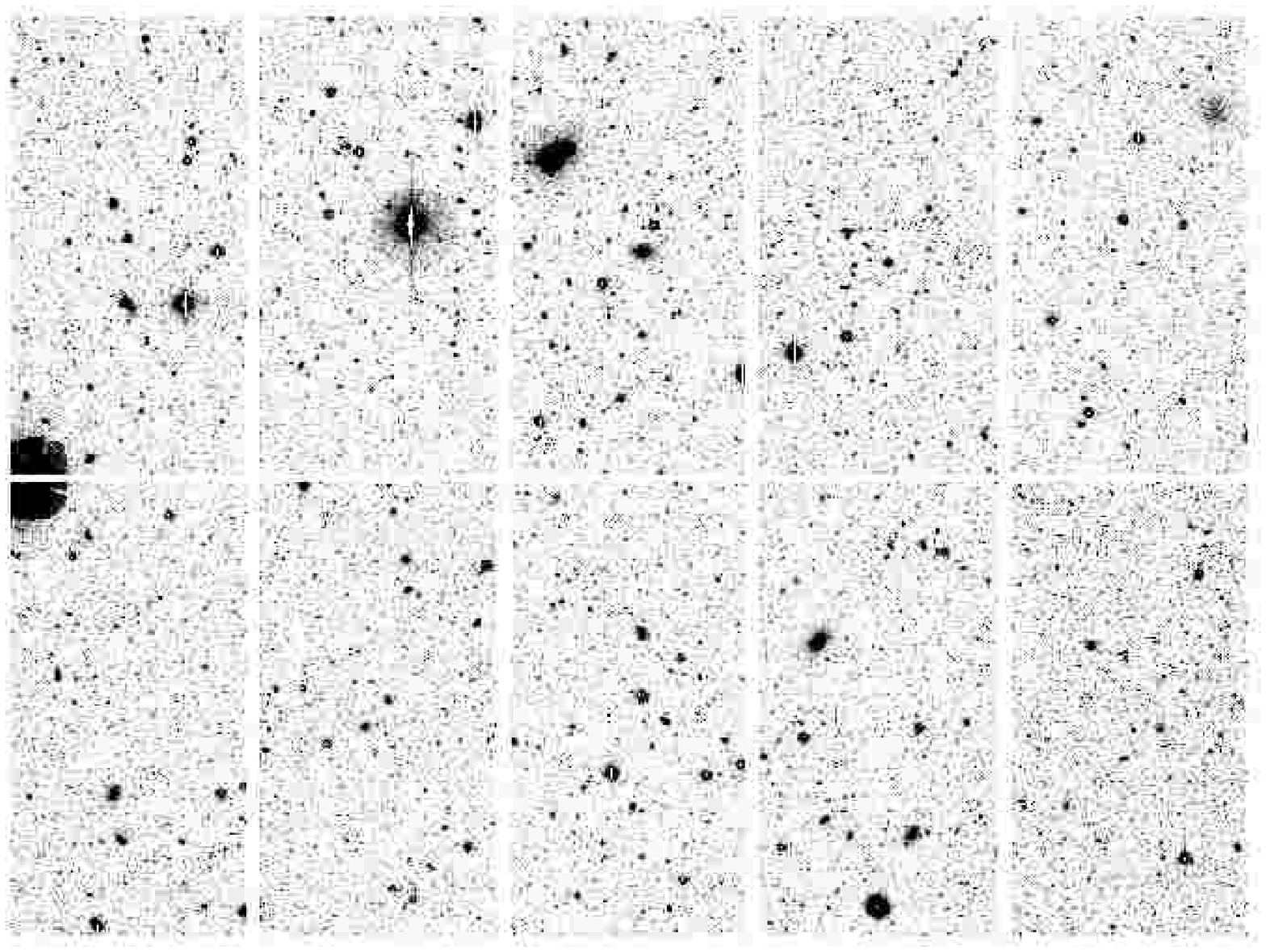}
 \end{center}
 \caption{Same as Fig. \ref{f1}, but for Field 2.} \label{f2}
\end{figure}

\begin{figure}
 \begin{center}
  \includegraphics[width=7.5cm,keepaspectratio]{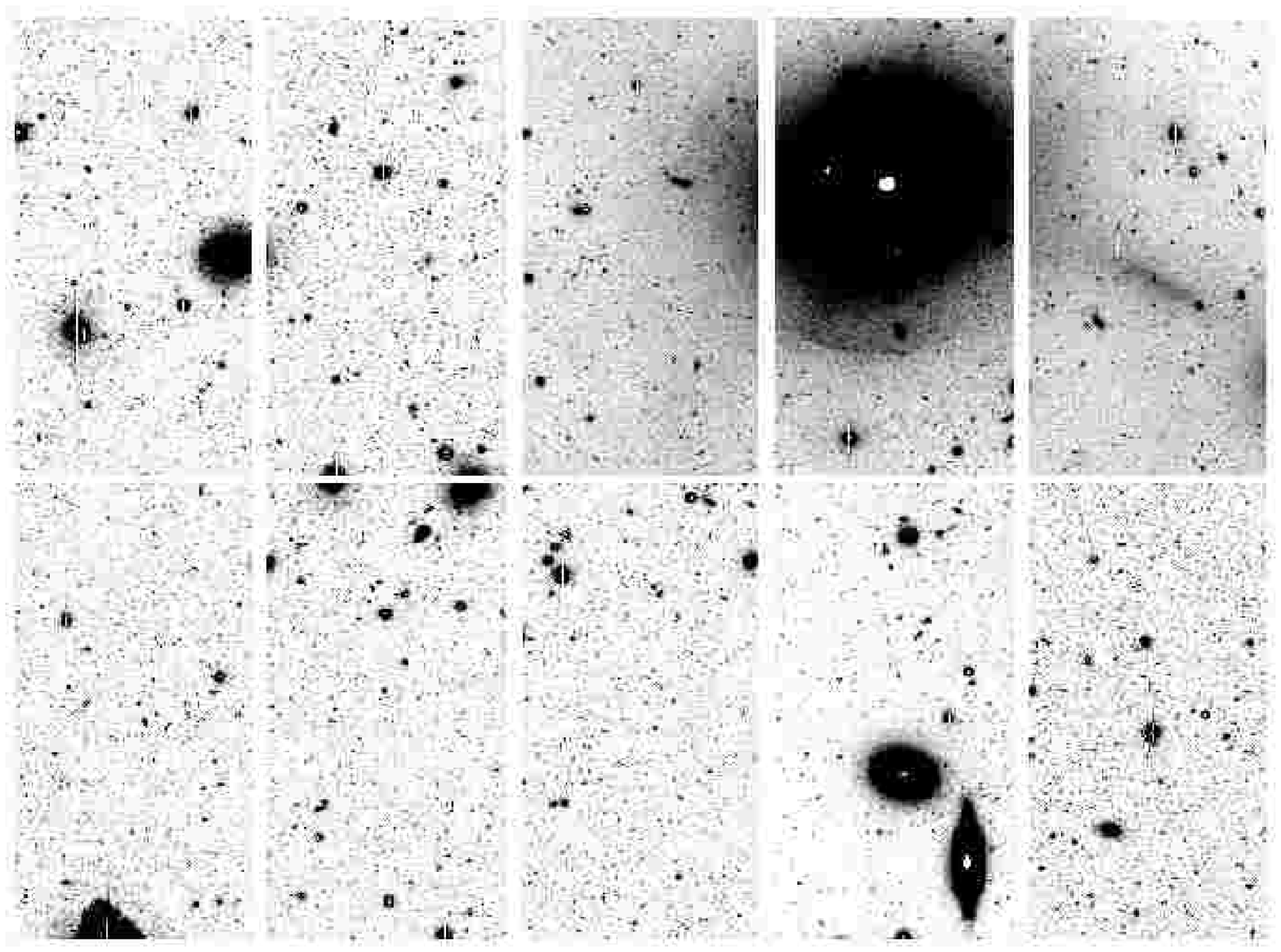}
 \end{center}
 \caption{Same as Fig. \ref{f1}, but for Field 3.} \label{f3}
\end{figure}

\begin{figure}
 \begin{center}
  \includegraphics[width=7.5cm,keepaspectratio]{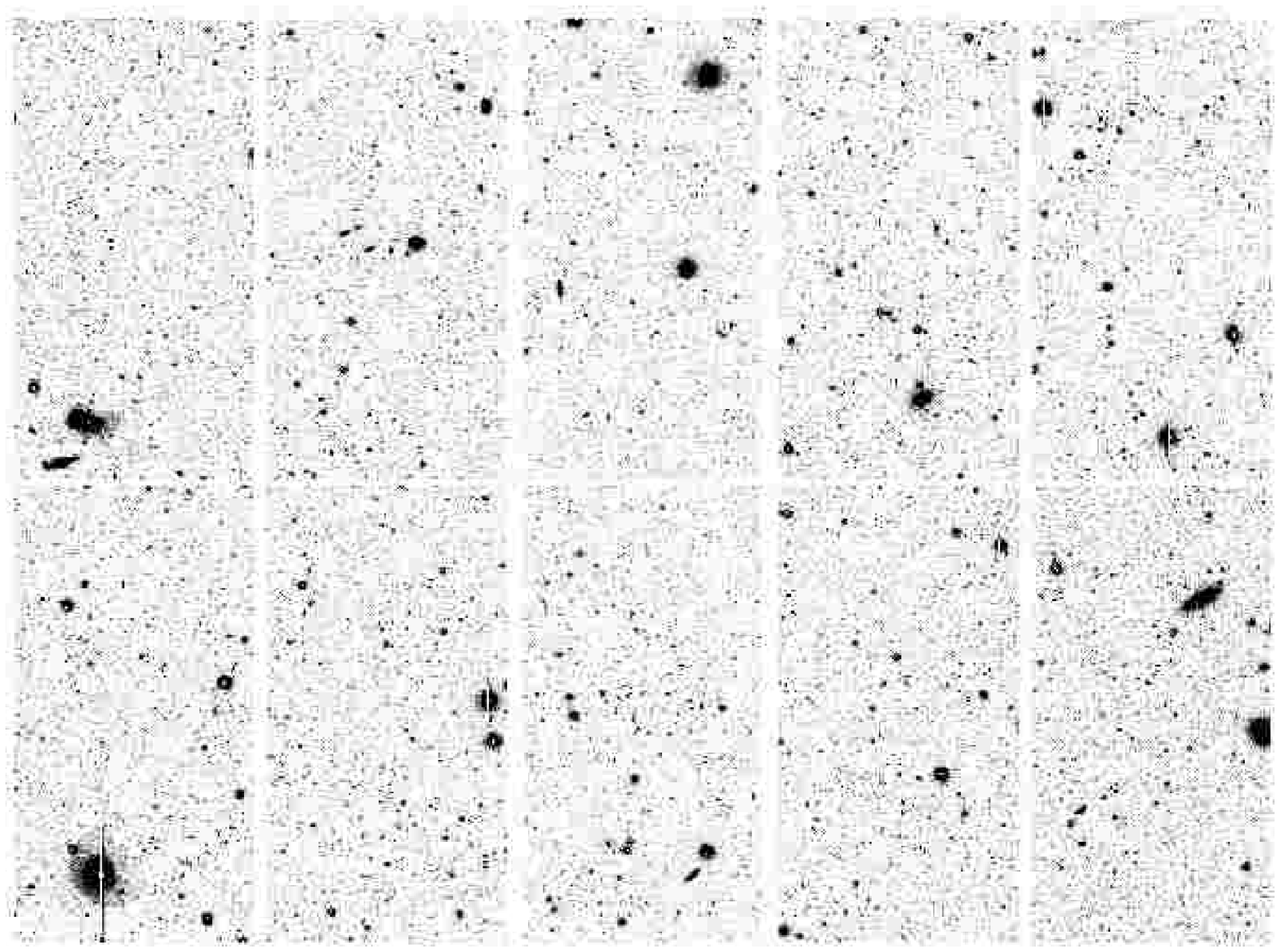}
 \end{center}
 \caption{Same as Fig. \ref{f1}, but for Field 4.} \label{f4}
\end{figure}

Data reduction was performed with IRAF\footnote{IRAF is distributed by
the National Optical Astronomy Observatories, which is operated by the
Association of Universities for Research in Astronomy, Inc. under
cooperative agreement with the National Science Foundation.} in a
standard manner; bias subtraction, flat-fielding, masking bad columns
and saturated pixels, sky subtraction, registration, and average
stacking with a 3 $\sigma$ clipping algorithm.
Sky subtraction was performed by employing the following two steps.
Firstly, an image was divided into a mesh of 128 $\times$ 128 pixels
($\sim 26^{\prime\prime} \times 26^{\prime\prime}$) and a median sky
value was estimated in each window after bright objects were masked. A
sky value in each pixel was then estimated by an interpolation of the
median sky values for the adjacent windows. A background image of an
object frame was created with this process which was then subtracted.
For CCD frames where bright galaxies or their envelopes are quite
extended (e.g., near M87 and NGC 4552), this method cannot be applied.
Instead, an average background was estimated as a single value using a
``blank'' CCD frame within the same exposure. It was corrected for the
sensitivity difference between the two CCD frames using the flat-field
frames.
In stacking CCD frames, aperture photometry of $20 - 30$ bright stellar
objects selected using SExtractor (Bertin \& Arnout 1996) based on the
\verb|CLASS_STAR| index was performed in each frame and the zeropoint of
the frame was shifted so as to match that in a frame taken on the second
night. PSF matching was not performed to avoid degradation of image
quality. The stacked images are presented in Fig. \ref{f1} $-$
\ref{f4}.

\subsubsection{Photometric and astrometric calibration}
\label{m87calib}

\begin{figure*}
 \begin{center}
  \includegraphics[width=8cm,angle=-90,keepaspectratio]{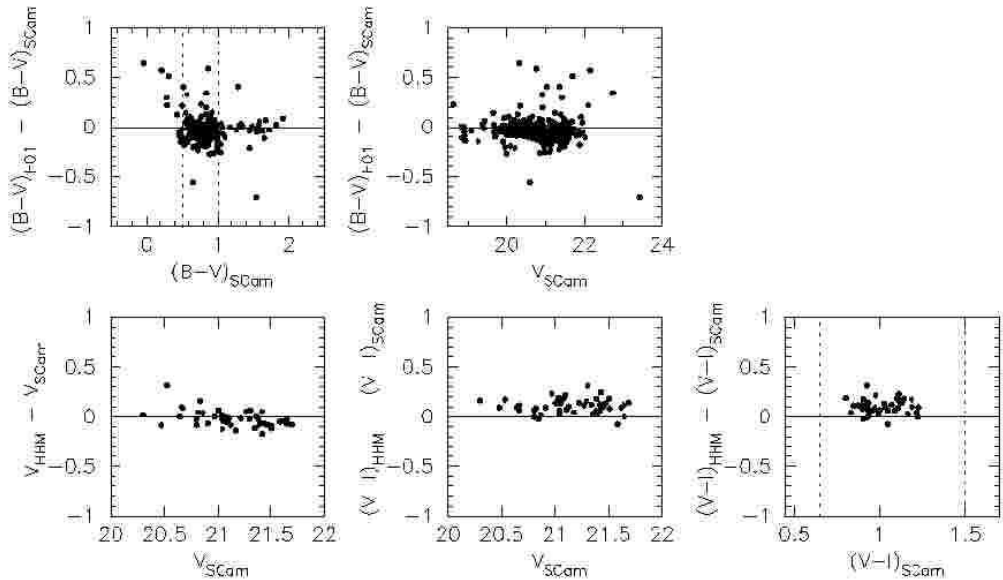}
  \caption{Comparison of our photometry ($V_{\rm SCam}$, $(B-V)_{\rm
  SCam}$, and $(V-I)_{\rm SCam}$) with that by H01 ($(B-V)_{\rm H01}$)
  and HHM ($V_{\rm HHM}$ and $(V-I)_{\rm HHM}$). Dotted lines in the
  plots for the colours approximately show reddest and bluest boundaries
  expected for GCs.} \label{h01hanes}
 \end{center}
\end{figure*}

Photometric calibration of the M87 fields was performed using standard
stars from Landolt (1992) which were observed at the beginning and end
of the second night. Several standard stars were imaged on each CCD chip
so that the calibration could be carried out individually. Photometry of
standard stars was performed within a $12^{\prime\prime}$ diameter
aperture. After excluding saturated stars and those in crowded regions,
magnitude zeropoints and the trends with sec $z$ and colours were
calculated. The estimated accuracy in the fitting procedure is $0.03 -
0.05$ mag.

Since our data include frames which were taken under non-photometric sky
conditions, we check the photometric calibration by comparing our
photometry of GCs around M87 with that in the literature. We make use of
GC photometry in the $V$ and $I$ bands by Harris, Harris, \& McLaughlin
(1998, HHM hereafter) and that in the Washington system ($C$ and $T_1$)
by Hanes et al. (2001, H01 hereafter). The GC photometry by H01 is
converted to the standard $BVI$ system by using the formulae obtained by
Geisler (1996).
PSF-fitting photometry was carried out for GCs using the PSF and ALLSTAR
tasks in the DAOPHOT package of IRAF. A PSF was determined using $\sim
20$ moderately bright (unsaturated) stellar objects in each CCD frame,
which were selected using SExtractor based on the \verb|CLASS_STAR|
index. The PSF obtained was also used for GC photometry, since the GCs
at the distance of the Virgo cluster are unresolved at the seeing sizes
of our images.
Galactic extinction was then corrected using reddening maps from
Schlegel, Finkbeiner, \& Davis (1998).
We are primarily concerned about zeropoint offsets for $V$-band
magnitude and $B-V$ and $V-I$ colours; the $V$ band image is used as a
selection band to make a catalog of GCs and $B-V$ and $V-I$ colours are
used to isolate GC candidates from other unresolved objects (see
\S~\ref{gcselection} for details). While $C-T_1$ can be converted into
$B-V$ with a small error, it is not converted to $V-I$ with a good
accuracy (Geisler 1996). We therefore decided to estimate the zeropoint
offset in $B-V$ from the comparison with H01 and those in $V$ and $V-I$
from the comparison with HHM. Our GC photometry is compared with that
from H01 and HHM in Fig. \ref{h01hanes}. Dotted lines in these plots
indicate the approximate edges of the colour range which is expected to
be occupied by GCs. These comparisons suggest that there are some
zeropoint offsets between our photometry and that in the literature. The
zeropoint offsets in $V$, $B-V$ and $V-I$ are 0.04 mag, 0.05 mag and
0.10 mag, respectively. Our magnitude and colours are corrected for these
zeropoint offsets in the following analyses.

We performed astrometric calibration against the 2MASS catalog using the
CCMAP task in IRAF. A plate solution (second order polynomial with full
cross terms) was computed for each CCD chip using stars with $J \leq 17$
mag. The fitting accuracy is typically $\sim 0_{\cdot}^{\prime\prime}2$.

\subsection{The Control Fields}\label{archive}

\begin{figure*}
 \begin{center}
  \includegraphics[height=12cm,keepaspectratio]{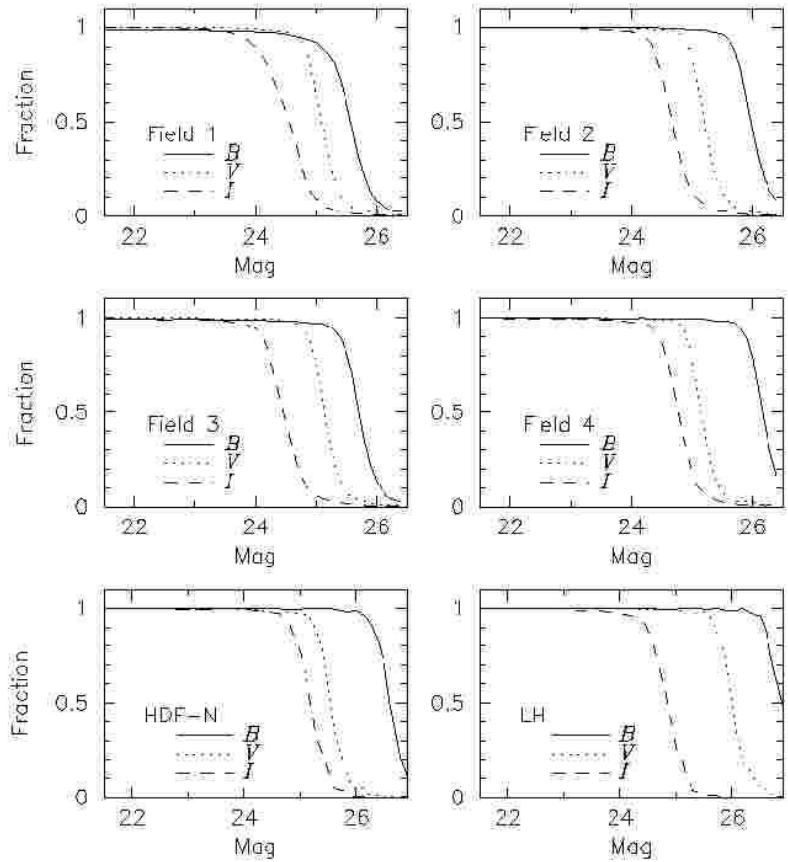}
 \end{center}
 \caption{Overall detection completenesses in $BVI$ bands are
 overplotted. Top and middle panels: the results in the M87 fields are
 shown. Bottom panels: those in the control fields are indicated.}
 \label{maglimbvi}
\end{figure*}

In addition to the M87 field data, we also analyze $BVI$ images of the
(blank) fields, HDF-N and Lockman Hole (LH), retrieved from the
Subaru-Mitaka-Okayama-Kiso Archive (SMOKA) system (Baba et al. 2002).
These data were taken with SCam on the Subaru telescope during several
observing runs in 2001: 23 and 24 Feb for $BVI$ on HDF-N and $B$ on LH,
and 22 and 23 Apr for $V$ and $I$ on LH (Capak et al.  2004). This
information is also summarized in Table \ref{basic}. These control field
data are used to estimate contamination in the sample of GC candidates
and to statistically subtract it, which is essential for investigating
properties of GC populations such as the luminosity function and colour
distribution. We emphasize that these control fields are also at high
galactic latitudes (Table \ref{basic}),
and that the data cover reasonably wide sky areas (one SCam field of
view: $\sim$ 900 arcmin$^2$) and are comparable with our data in the M87
field in terms of the filter set, limiting magnitudes, and image
quality. This minimizes the possibility of introducing any systematic
errors into the subtractive corrections for foreground and background
contamination in the GC sample.

The data reduction was carried out with SDFRED (Yagi et al. 2002; Ouchi
et al. 2004), which is a reduction pipeline optimized for SCam data of
blank fields. The basic reduction procedure is the same as that applied
to the M87 field data. The large scale of telescope dithering ($\sim
1^{\prime}$) for these control field data enables the CCD frames to be
stacked into one continuous image with sensitivity differences between
CCDs corrected by using stellar objects in the overlap regions.
To avoid any complications due to a possible drop of limiting magnitude
near the field edge, we use only the central $27^{\prime} \times
27^{\prime}$ region in the following analyses.
The magnitude zeropoints of the HDF-N data were calculated using the
photometry catalog by Capak et al. (2004). These authors did not observe
any standard stars during the observations and determined magnitude
zeropoints by exploiting the accurate photometry of objects in the
region where deep HST/WFPC2 data are available. The best-fit SEDs to the
multi-band photometry of the objects (Fern\'{a}ndez-Soto, Lanzetta \&
Yahil 1999) were used to account for the slight differences in the
filter responses between Subaru/SCam and HST/WFPC2. Since the $B$ band
data of the LH field were taken in the same runs as for the HDF-N, the
same zeropoint is adopted. The $V$ and $I$ band data of the LH field
were taken on different observing runs and standard stars were observed
at elevations similar to those of the LH field during the night.
Magnitude zeropoints were derived from these data.
Galactic extinction was then corrected using Schlegel et al. (1998).

\section{DATA ANALYSES}\label{analyses}

\subsection{Halo Light Subtraction}\label{subtraction}

As shown in Figs. \ref{f1} and \ref{f3}, M87 and NGC 4552 extend
across several CCD frames and their halos have to be subtracted to
reveal the GC populations. We removed the halos by conducting an
iterative median smoothing and subtraction (e.g., McLaughlin, Harris, \&
Hanes 1994). First, we subtract unresolved objects from an image. This
process is not mandatory, but it helps better model the extended halo
light distributions or their residuals in subsequent iterations. Object
detection was performed with SExtractor and unresolved objects were
picked out based on the \verb|CLASS_STAR| index. A PSF was determined
using moderately bright stars, which was then fitted to unresolved
objects and the fitted profiles were subtracted from the original image
using the ALLSTAR task. The resulting image was median-smoothed
to create an image with a halo light distribution which was subtracted
from the original image. This procedure was repeated 4 times with the
mesh size successively reduced (128, 64, 32, and 32 pixels).

\begin{figure*}
 \begin{center}
  \includegraphics[height=12cm,keepaspectratio]{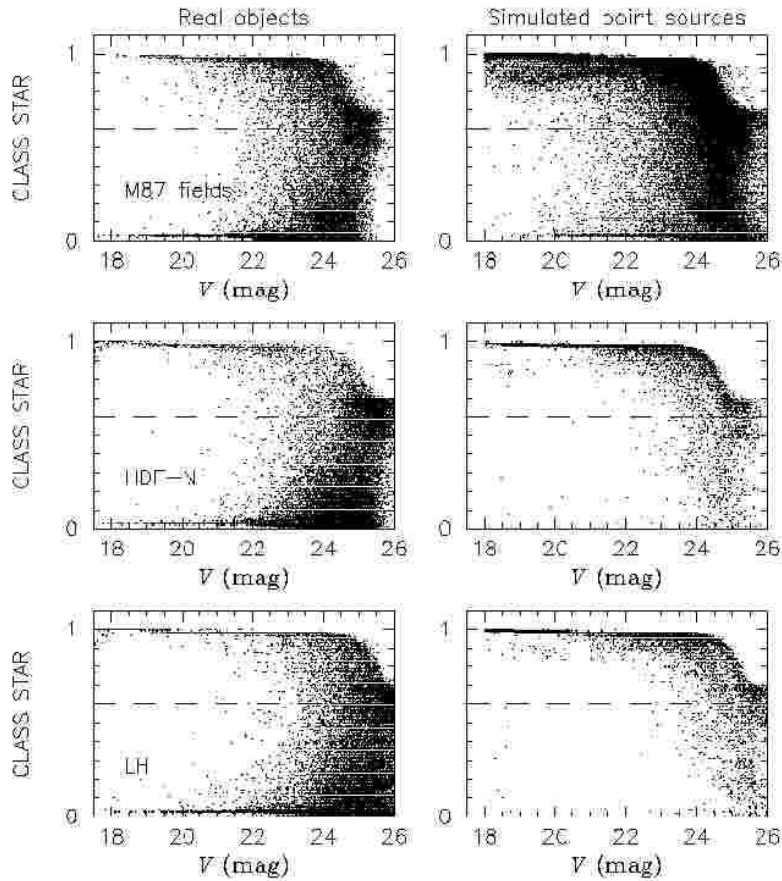}
 \end{center}
 \caption{\texttt{CLASS\_STAR} indices of detected objects in the $V$
 band images of the M87 fields, HDF-N field, and LH field are plotted
 against their magnitudes in the top, middle, and bottom panels,
 respectively. Dashed line indicates the lower limit for selection of
 unresolved objects (0.6).}
 \label{classstar}
\end{figure*}

\subsection{Artificial Star Tests}\label{limitmag}

\begin{figure*}
 \begin{center}
  \includegraphics[height=14cm,angle=-90,keepaspectratio]{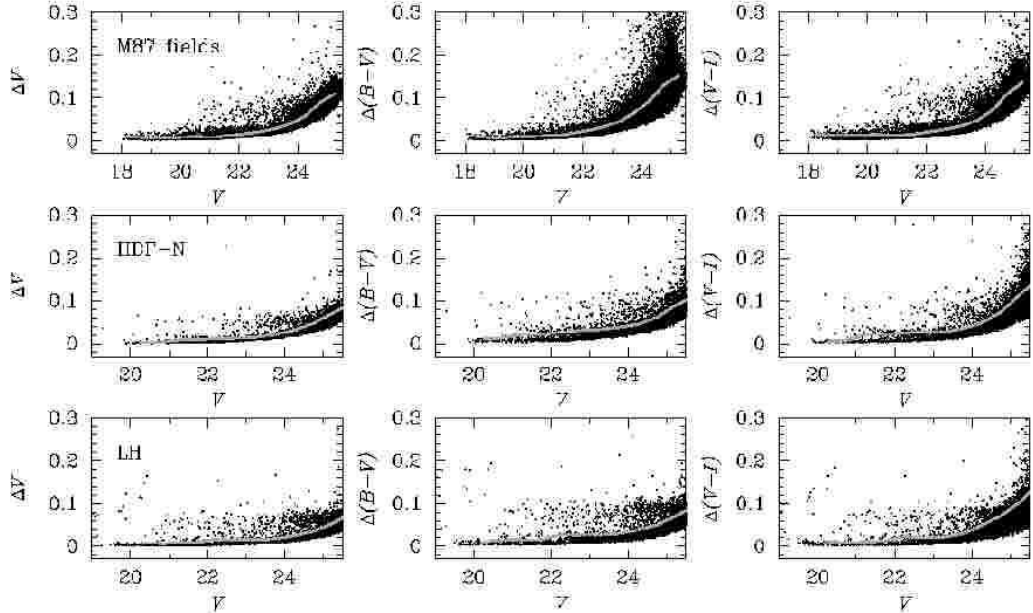}
 \end{center}
 \caption{For the unresolved objects, errors of $V$-band magnitude
 (left), $B-V$ colour (centre) and $V-I$ colour (right) from the
 PSF-fitting photometry are plotted as a function of $V$-band
 magnitude. Grey line indicates the boundary below which 80 \% of the
 unresolved objects are included at a given magnitude. In the top,
 middle, and bottom panels, the results in the M87 fields, HDF-N field,
 and LH field are shown, respectively.} \label{magerrs}
\end{figure*}

\begin{figure*}
 \begin{center}
  \includegraphics[height=12cm,angle=-90,keepaspectratio]{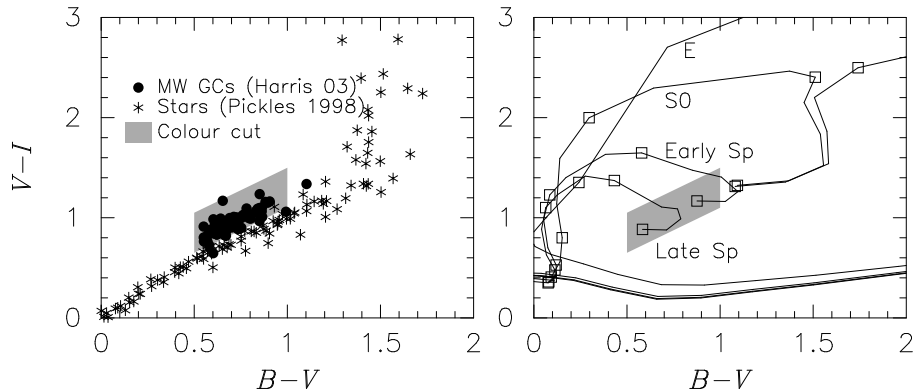}
  \caption{The colour criterion to select GC candidates is indicated by
  a shaded region. In the left panel, colours of Galactic GCs in the
  catalog by Harris (2003 version) are plotted with circles, and stellar
  colours based on a stellar flux library by Pickles (1998) are plotted
  with asterisks. Galactic extinctions for the GC colours are corrected
  using the colour excess of each GC in the catalog. In the right panel,
  evolutionary tracks of the galaxy colours calculated using PEGASE v2.0
  (Fioc \& Rocca-Volmerange 1997) are indicated. Open squares are
  plotted at redshifts of 0, 0.5, 1.0, 1.5, and 2.0 on each track. Note
  that the galaxy colours at $z = 0$ are plotted around the middle of
  this panel and go towards outer regions at higher redshifts.}
  \label{colourcut}
 \end{center}
\end{figure*}

We performed artificial star tests for investigating detection
completeness to point sources on the $B$-, $V$-, and $I$-band images,
after bright galaxies were subtracted if necessary. Using the IRAF
STARLIST and MKOBJECT tasks, artificial stars with the same PSF as that
determined using real unresolved objects were distributed on the
original image. For a series of tests, 500 artificial stars within a
certain range of magnitude $[m, m + 0.5$ mag$]$ were generated while
successively changing the magnitude range as the test progressed.
SExtractor was then used for object detection and the fraction of
artificial stars detected (i.e., detection completeness) was calculated
as a function of magnitude. Note that faint stars could be rejected by
DAOPHOT even if they are detected by SExtractor, but the fraction of
such artificial stars turns out to be small ($\leq$ 1 \%) throughout the
magnitude range investigated in our artificial star tests. We performed
these artificial star tests on all the CCD frames in all the observed
fields (Field 1 $-$ 4)
and the overall completeness in each observing field is plotted against
$B$-, $V$-, and $I$-band magnitude in the top and middle panels of
Fig. \ref{maglimbvi}. This indicates the presence of a slight
field-to-field variation in limiting magnitude.
The detection completeness is also a function of galactocentric
distance; for instance, limiting magnitudes are $\sim$ 0.5 mag brighter
near the centre of M87 where the noise is higher. Therefore, when
investigating the luminosity function and colour distribution of GC
candidates within an annulus at a certain distance from the host galaxy,
we correct for incompleteness using the completeness functions estimated
within the same annulus. Especially near the luminous ellipticals, we
divide the annulus into sub-annuli to follow the local variation of the
incompleteness in the annulus. The artificial star tests were also
executed on the control field images and the overall completeness
functions are indicated in the bottom panels of Fig. \ref{maglimbvi}.
The magnitudes giving 50 \% completeness on the M87 fields and the
control fields are listed in Table \ref{basic}.

Since unresolved objects are firstly selected based on the
\verb|CLASS_STAR| indices on the $V$-band image when selecting GC
candidates (see \S~\ref{gcselection}), one also needs to consider biases
associated with the selection of unresolved objects in addition to the
simple detection incompleteness as mentioned above. In the top panels of
Fig.  \ref{classstar}, \verb|CLASS_STAR| indices of the detected objects
(left panel) and the artificial stars (right panel) in the M87 fields
are plotted against $V$-band magnitude. In the middle and bottom panels,
the results in the control fields are shown. These plots indicate that
the \verb|CLASS_STAR| index of a point source tends to be underestimated
at fainter magnitudes and more stellar objects are expected to be
excluded when we classify objects with \verb|CLASS_STAR| indices larger
than a certain value as unresolved. We quantify this selection effect as
functions of $V$ magnitude and distance from the host galaxy based on
these artificial star tests.

\subsection{Selection and Photometry of GC Candidates}
\label{gcselection}

\begin{figure*}
 \begin{center}
  \includegraphics[height=17.5cm,keepaspectratio]{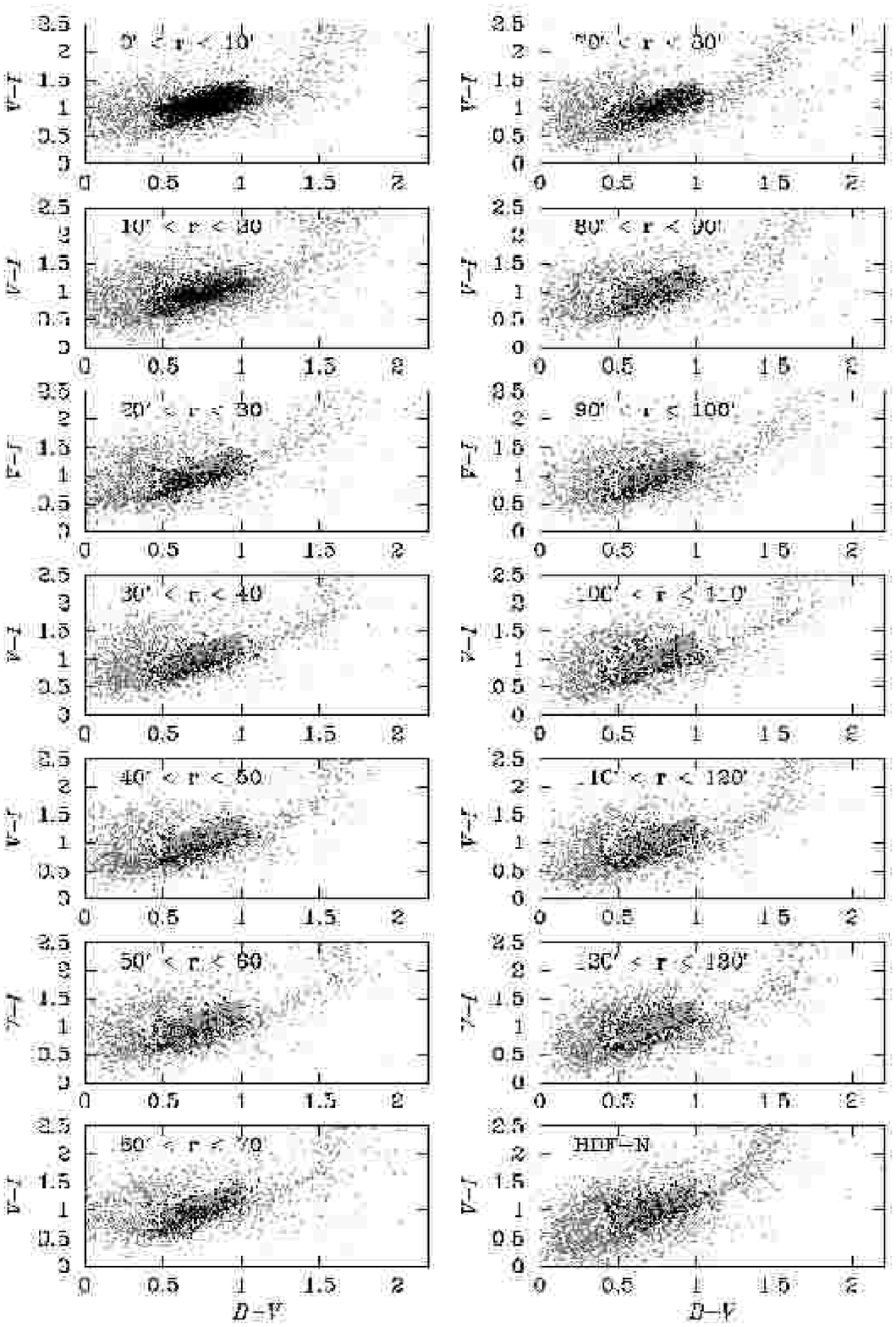}
 \end{center}
 \caption{$B-V$ and $V-I$ colour-colour diagrams for the unresolved
 objects. The data in the M87 fields are divided into panels based on
 the distances of the sources from M87. In the bottom right panel, the
 unresolved objects in the HDF-N data (see \S~\ref{archive}) are
 plotted. Black dots indicate unresolved objects which pass the colour
 cut, and grey dots are those which do not. The shaded region in each
 panel indicates the colour criterion. Some objects sitting outside the
 colour criterion are accepted because we take into account errors in
 both of the colours in applying the colour cut to these objects (see
 text for details).} \label{allbvi}
\end{figure*}

We begin selection of GC candidates with object detection using
SExtractor. Firstly, we picked out all objects having at least 20
connected pixels ($\sim 0.8$ arcsec$^2$, which is approximately equal to
the FWHM area of the PSF) more than 2 $\sigma$ above the local
background (only objects selected in all the three $B$-, $V$-, and
$I$-bands were used for analysis). Secondly, we selected objects with
\verb|CLASS_STAR| indices larger than 0.6 as unresolved objects. The
$V$-band image was used for this classification because it has the best
image quality in our data. The subsequent results do not change if this
cutoff for the \verb|CLASS_STAR| index is set to 0.5 or 0.7. We note
that our criterion (\verb|CLASS_STAR| $\geq 0.6$) is more stringent than
those adopted in previous GC studies ($\geq 0.4$ in Dirsch et al. 2003
and $\geq 0.35$ in Forbes et al. 2004) using data with better image
quality than our data.

PSF-fitting photometry was performed for these unresolved objects and
their $B$, $V$, and $I$ magnitudes were measured. In this process, a
residual sky background around an object is estimated within an annulus
10$^{\prime\prime}$ away from the object with a 4$^{\prime\prime}$ width
and is subtracted. Errors in the $V$-band magnitude and $B-V$ and $V-I$
colours due to a PSF fitting error and sky subtraction error are plotted
as a function of $V$-band magnitude in the top, middle, and bottom
panels of Fig. \ref{magerrs} for unresolved objects in the M87 fields,
HDF-N field, and LH field, respectively. A colour criterion was then
imposed on these unresolved objects to isolate GC candidates. We show
this colour criterion as the shaded region in Fig.  \ref{colourcut}
which includes almost all the Galactic GCs in the catalog by Harris
(1996)\footnote{The catalog used here is the version last updated in
2003.} but minimizes the contamination by foreground stars and
background galaxies. The $B-V$ and $V-I$ colour-colour diagrams of the
unresolved objects on our images are indicated in Fig. \ref{allbvi}; the
objects are divided into panels based on their distances from M87, apart
from those in the HDF-N field, which are plotted in the bottom right
panel for reference. Black dots indicate unresolved objects which pass
the colour selection, while grey dots are those which do not satisfy the
colour criterion. Some unresolved objects sitting outside the colour
criterion are accepted as GC candidates by taking into account errors in
the colours (e.g., Rhode \& Zepf 2001); if colours of unresolved objects
can satisfy the colour criterion within their errors, they are sampled
as GC candidates. This ``inclusive'' colour selection allows us to
incorporate fainter GCs which are more likely to be scattered out of the
colour criterion due to larger errors in the colours. Although
contaminating objects may also be included, they are expected to be
corrected for by the control field data (see \S~\ref{fieldpop}).

\subsection{Correction of Incompleteness and Foreground and Background
Contamination} \label{fieldpop}

\begin{figure*}
 \begin{center}
  \includegraphics[height=12cm,angle=-90,keepaspectratio]{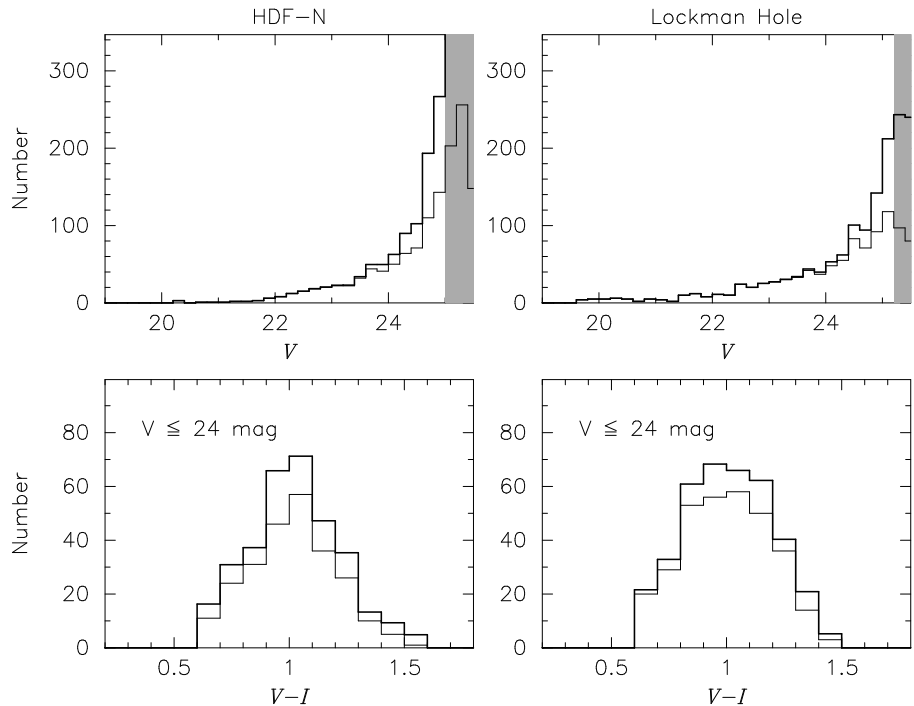}
 \end{center}
 \caption{{\it Upper panels}: Luminosity functions of the unresolved
 objects which pass the colour selection in the HDF-N ({\it left}) and
 the LH ({\it right}). Thin line shows raw LF and thick line indicates
 incompleteness-corrected LF. Shaded region at the right hand side
 corresponds to the magnitude range where the completeness is lower than
 50 \%. {\it Lower panels}: Colour distributions of the unresolved
 objects ($V \leq 24$ mag) which pass the colour selection. Thin line
 shows raw distribution and thick line indicates
 incompleteness-corrected distribution.}  \label{fieldlfcld}
\end{figure*}

\begin{figure*}
 \begin{center}
  \includegraphics[height=17cm,keepaspectratio]{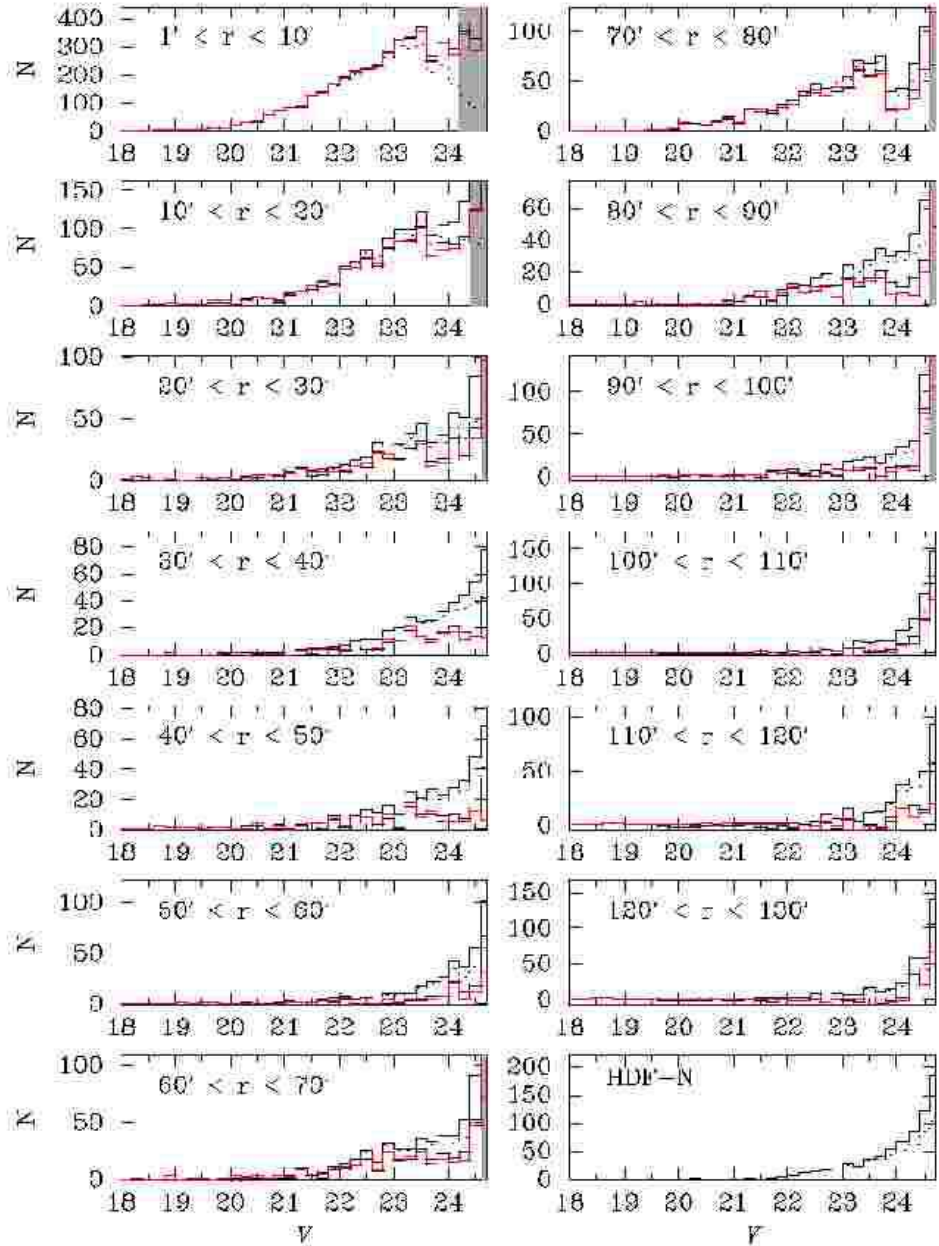}
 \end{center}
 \caption{GC Luminosity Functions (GCLFs) obtained within annuli
 centered on M87 are indicated. The dotted line indicates a raw LF
 (without incompleteness correction or control field subtraction) and
 thin black line describes an LF after the incompleteness is corrected
 (but no field subtraction is performed yet). The red line and the thick
 black line are those where the control field subtraction is performed
 after the incompleteness correction: red (black) line shows the LF
 where the control field subtraction is performed using the HDF-N (LH)
 data, respectively. The shaded region indicates the magnitude range
 where the completeness is lower than 50 \%. In the bottom right panel,
 the LF of unresolved and colour-selected contaminating objects in the
 HDF-N field is shown for reference (survey area is not normalized). }
 \label{gclf_cor}
\end{figure*}

\begin{figure*}
 \begin{center}
  \includegraphics[height=17cm,keepaspectratio]{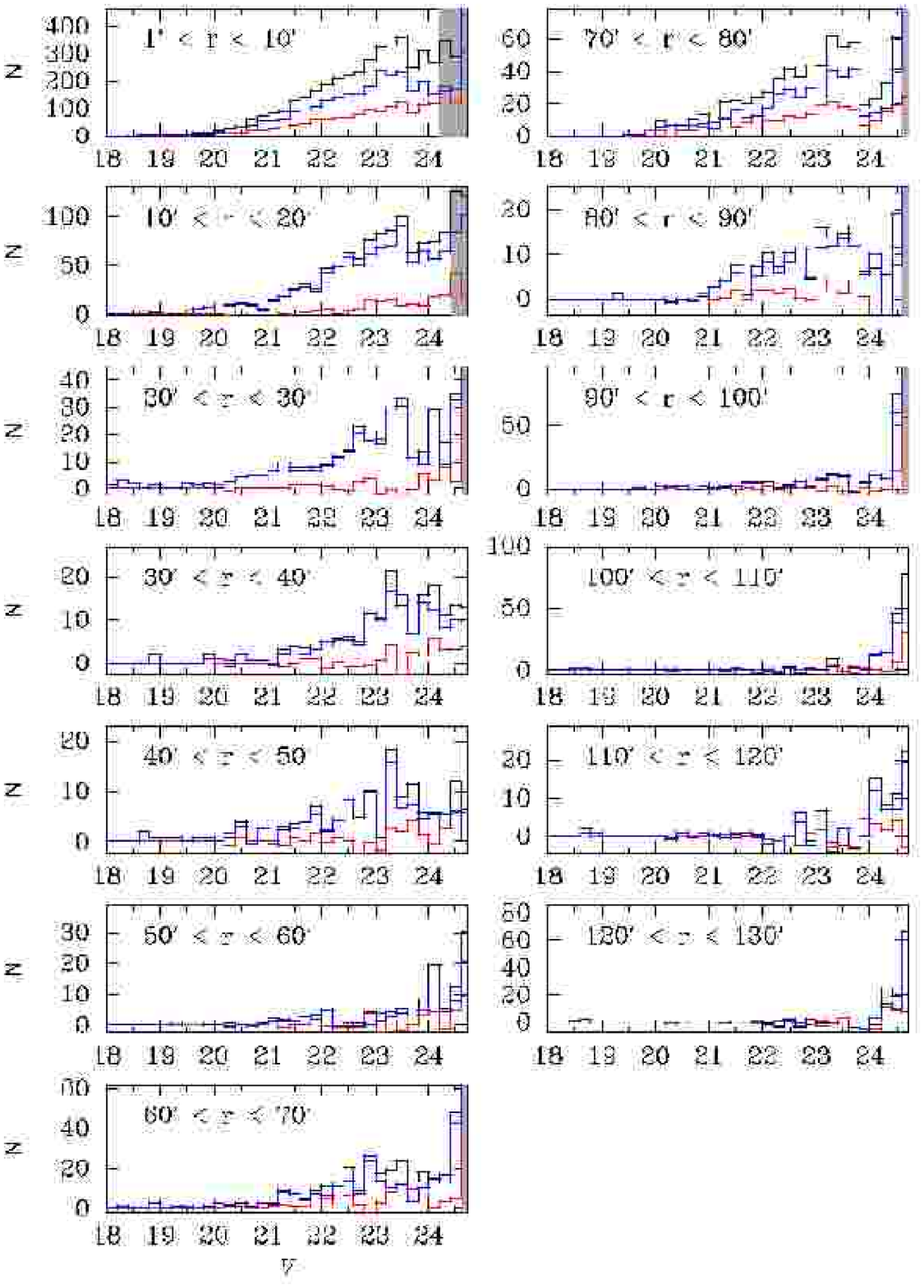}
 \end{center}
 \caption{Black line indicates GCLF after the incompleteness is
 corrected and the control field is subtracted using the HDF-N data
 (same as that described with the red line in Fig. \ref{gclf_cor}).
 This is divided into two LFs depending on GC colour: Red and blue lines
 indicate the LFs for red ($V-I > 1.1$) and blue ($V-I \leq 1.1$) GC
 candidates, respectively.} \label{gclf}
\end{figure*}

In deriving a GC luminosity function (GCLF) and colour distribution,
incompleteness correction is undertaken as follows. The number of GC
candidates at a certain $V$ magnitude is firstly corrected for the
incompleteness in detection and selection of unresolved objects on the
$V$-band image. Detection incompleteness on the $B$ and $I$ band images
is then corrected; GCs with a given $V$-band magnitude are divided into
bins according to their $B-V$ and $V-I$ colours, and the numbers of
objects are multiplied by a factor to correct for the incompleteness at
the $B$ and $I$ magnitudes ($B = (B-V) + V$, $I = V - (V-I)$). These
incompleteness corrections are also applied to GC candidates found in
the control fields.

Although the selection using the two colour diagram is expected to
efficiently isolate GCs from foreground stars and background unresolved
galaxies, there are still likely to be some contaminating objects, and a
subtractive correction of this contamination is essential to investigate
GC properties in the outer halo of the host galaxy where the GC surface
number density is very low. We extract contaminating populations of
unresolved objects from the control fields by using selection criteria
identical to those adopted in the M87 fields; objects with
\verb|CLASS_STAR| $\geq$ 0.6 are selected as unresolved objects and GC
candidates are isolated by using the same colour criterion on the $B-V$
vs. $V-I$ colour-colour diagram.

When subtracting contamination in a certain region within the M87 survey
field, we take into account differences in data quality, especially
errors in the colours at a certain $V$ magnitude, between the M87 field
and control field. Since the errors in the control fields are smaller
than those in the M87 field, a smaller number of objects would be
scattered into the colour criterion in the control fields and the
contamination could be underestimated. We pick out unresolved objects
within narrow ranges ($\pm 0.1$ mag) of $V$, $B-V$ and $V-I$ from both
the M87 field and control field and calculate the differences of typical
errors in the colours between the two samples of unresolved objects. We
then randomize the measured colours of the unresolved objects in the
control field by an amount which is determined from a Gaussian
distribution whose average is zero and whose standard deviation is
estimated from the difference of the typical errors in the colours. The
observational errors in the colours of the unresolved objects in the
control field are therefore replaced with the typical errors of those in
the M87 field. This sequence is repeated in the control field for
different $V$ magnitudes and the colours successively changed, which
provides a mock catalog of unresolved objects in the control field whose
error characteristics are compatible with those in the M87 field. Based
on the colours and errors in this mock catalog, the colour selection is
performed to pick up contaminating objects in the control
field\footnote{Because this sequence involves random numbers, the GC
candidates in a control field needs to be defined by a number of
attempts based on the Monte Carlo technique. But in fact, the variance
of the average population is small because the ``inclusive'' selection
of GC candidates does not give a sharp cutoff on the $B-V$ and $V-I$
colour-colour diagram.}. The luminosity function and colour distribution
of these contaminating objects including the incompleteness corrections
are then subtracted from those in the M87 field with the survey area
normalized.

In Fig. \ref{fieldlfcld}, the $V$-band luminosity functions and $V-I$
colour distributions of GC candidates found in the control fields are
displayed. Note that these are obtained by considering the entire region
of the M87 survey field. The LF and colour distribution in the control
fields are normalized to the survey area of a target field when
subtracted. Most of these contaminating objects are likely to be
background galaxies with compact morphology; candidates are dwarf
ellipticals and blue compact dwarfs, the latter of which need to be at
higher redshifts than the former to meet the colour criterion.
Foreground stars can also be scattered into the GC colour selection due
mainly to photometric errors, but since the errors in the colours are
$\sim 0.1$ mag even at $V \sim 24.5$ mag, which is approximately the
faintest magnitude of GCs studied in this work, their contribution is
presumed to be smaller than the background galaxies.

\setlength{\tabcolsep}{1mm}
\begin{table*}
\centering
\begin{minipage}{140mm}
\begin{tabular}{llcccccc} \hline\hline
 \multicolumn{2}{c}{Galaxy} &
 \multicolumn{2}{c}{All GCs} &
 \multicolumn{2}{c}{Red GCs} &
 \multicolumn{2}{c}{Blue GCs} \\ \cline{3-4}\cline{5-6}\cline{7-8}
  &
  &
 $V_{\rm TO}$ &
 $\sigma$ &
 $V_{\rm TO}$ &
 $\sigma$ &
 $V_{\rm TO}$ &
 $\sigma$ \\
  &
  &
 (mag) &
 (mag) &
 (mag) &
 (mag) &
 (mag) &
 (mag) \\ \hline

 M87      & ($1^{\prime} \leq R \leq 10^{\prime}$) &
 $23.62 \pm 0.06$ & $1.50 \pm 0.04$ & $23.85 \pm 0.19$ & $1.57 \pm 0.09$ & $23.35 \pm 0.05$ & $1.38 \pm 0.04$ \\
 M87      & ($1^{\prime} \leq R \leq 5^{\prime}$)  &
 $23.63 \pm 0.08$ & $1.48 \pm 0.05$ & $23.77 \pm 0.23$ & $1.54 \pm 0.11$ & $23.36 \pm 0.11$ & $1.44 \pm 0.05$ \\
 M87      & ($5^{\prime} \leq R \leq 10^{\prime}$) &
 $23.52 \pm 0.08$ & $1.44 \pm 0.05$ & $23.97 \pm 0.43$ & $1.48 \pm 0.22$ & $23.37 \pm 0.08$ & $1.36 \pm 0.05$ \\
 M87      & ($R \lesssim 1^{\prime}$; K99) &
 $23.67 \pm 0.07$ & $1.39 \pm 0.06$ & $-$ & $-$ & $-$ & $-$  \\
 M87      & ($R \lesssim 1^{\prime}$; L01) & 
 $23.44^{+0.04}_{-0.08}$ & $-$ & $23.52^{+0.06}_{-0.08}$ & $-$ & $23.30^{+0.06}_{-0.12}$ & $-$ \\
 NGC 4552 & ($1^{\prime} \leq R \leq 10^{\prime}$) &
 $23.56 \pm 0.20$ & $1.34 \pm 0.12$ & $23.75 \pm 0.35$ & $1.40 \pm 0.18$ & $23.33 \pm 0.16$ & $1.09 \pm 0.10$ \\
 NGC 4552 & ($1^{\prime} \leq R \leq 5^{\prime}$) &
 $23.48 \pm 0.22$ & $1.42 \pm 0.14$ & $23.85 \pm 0.44$ & $1.59 \pm 0.24$ & $23.52 \pm 0.28$ & $1.31 \pm 0.18$ \\
 NGC 4552 & ($R \lesssim 1^{\prime}$; KW01) & 
 $23.54 \pm 0.18$ & $1.3$ (fixed) & $-$ & $-$ & $-$ & $-$ \\
 NGC 4552 & ($R \lesssim 1^{\prime}$; L01) & 
 $23.19^{+0.11}_{-0.15}$ & $-$ & $23.52^{+0.22}_{-0.20}$ & $-$ & $22.91^{+0.15}_{-0.18}$ & $-$ \\ \hline

\end{tabular}

\caption{Parameters of Gaussians fitted to GCLFs. In the outer region of
 NGC 4552, no Gaussian fits were attempted due to the poor statistics.
 The GCLF parameters in the core region ($R \lesssim 1^{\prime}$) of M87
 and NGC 4552 taken from the literature are also shown for comparison:
 Kundu et al. (1999; K99), Kundu \& Whitmore (2001; KW01), and Larsen et
 al. (2001; L01). L01 fitted $t_5$ functions to the GCLFs and hence we
 only indicate the $V_{\rm TO}$s.}  \label{gclfparams}

\end{minipage}
\end{table*}

\begin{figure*}
 \begin{center}
  \includegraphics[height=14cm,angle=-90,keepaspectratio]{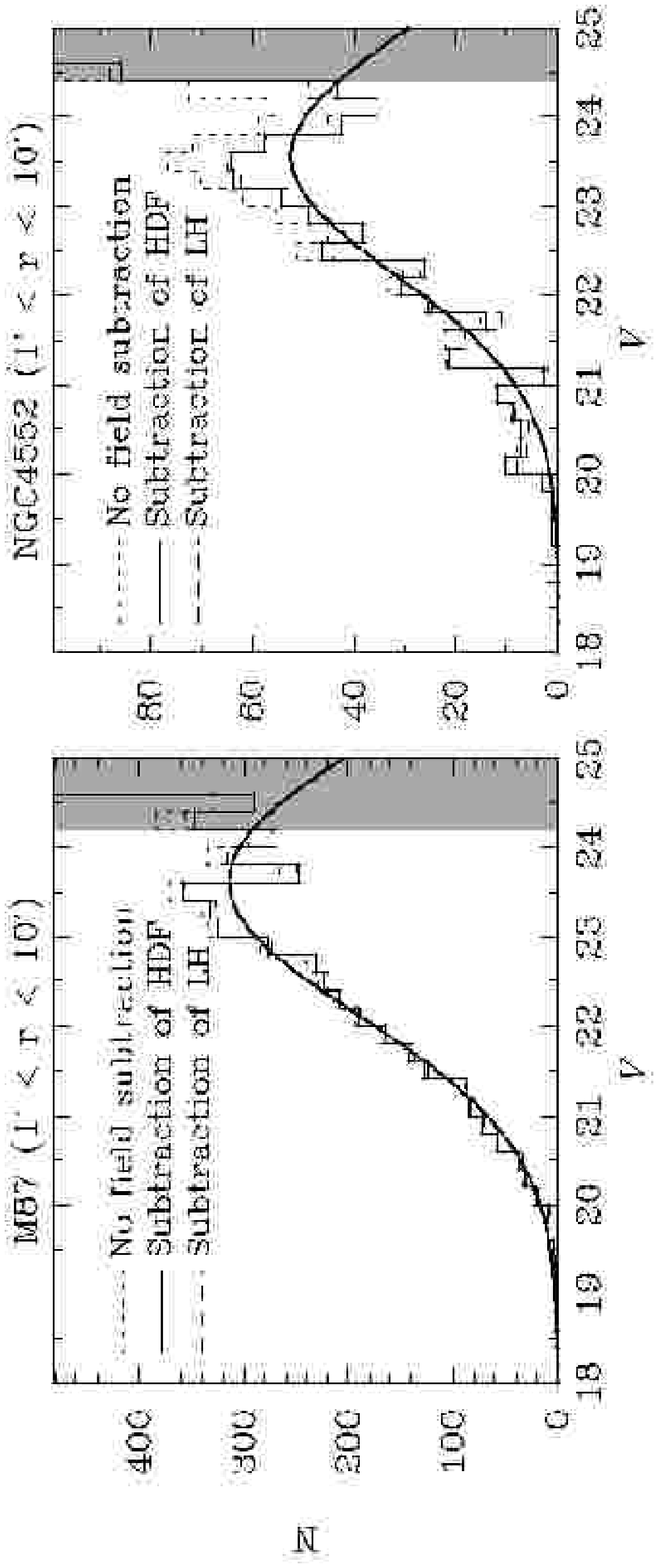}
 \end{center}
 \caption{Histograms show GCLFs obtained only using GCs at distances
 $\leq 10^{\prime}$ from host galaxy centre. Dotted line indicates
 incompleteness-corrected GCLF. Solid and dashed lines show GCLFs after
 subtracting control field populations based on the HDF-N and LH data,
 respectively. A Gaussian fitted to the GCLF is overplotted by a solid
 line. Shaded region indicates the magnitude range where the
 completeness is lower than 50 \%.}  \label{gauss}
\end{figure*}

\section{RESULTS AND DISCUSSIONS}
\label{resultsanddiscussions}

\subsection{Globular Cluster Luminosity Function}

In Fig. \ref{gclf_cor}, GCLFs obtained within annuli centered on M87 are
presented. The dotted lines indicate raw GCLFs without incompleteness
correction or control field subtraction, and the thin black lines show
GCLFs after the incompleteness is corrected for (but no subtractive
correction for contamination is performed yet). The red (thick black)
lines show GCLFs after the control field subtraction is also performed
based on the HDF-N (LH) field, respectively.
In Fig. \ref{gclf}, the GCLFs after the incompleteness correction and
field subtraction using the HDF-N data are divided into red ($V-I >
1.1$) and blue ($V-I \leq 1.1$) GC subpopulations. The black line shows
the GCLF for the total GC population (blue $+$ red). The boundary colour
of the blue and red GCs corresponds approximately to the middle of the
peak colours in the bimodal colour distributions (see Paper II). Note
that there is a local maximum of GC number in the $70^{\prime} < r <
80^{\prime}$ bin due to the GC population of NGC 4552 at a distance of
$\sim$ 75$^{\prime}$ from M87. GCLFs in the outermost regions tend to
have some negative bins due to the field subtraction. In both of these
figures, the shaded region indicates the magnitude range where the
completeness is lower than 50 \%. Note that the completeness is
calculated by considering not only the simple detection completeness on
the $V$ band image but also the selection efficiency of unresolved
objects based on the \verb|CLASS_STAR| index and the completeness in $B$
and $I$ bands at the corresponding magnitudes depending on the GC
colours.
We investigate and discuss the spatial distributions of GC populations
in detail in Paper II.

\begin{figure*}
 \begin{center}
  \includegraphics[height=14cm,angle=-90,keepaspectratio]{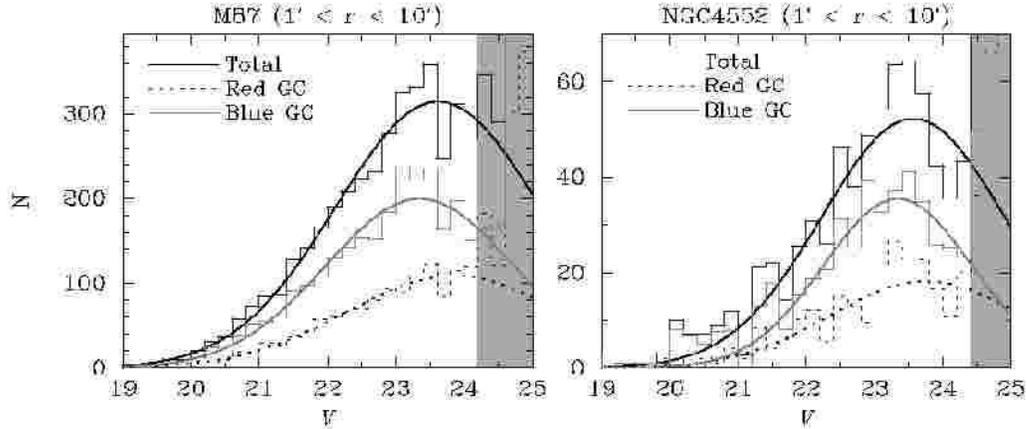}
 \end{center}
 \caption{The GCLF of all GCs, red GCs, and blue GCs is indicated by a
 black solid line, dotted line, and grey solid line, respectively.
 Incompleteness was corrected and contamination was subtracted based on
 the HDF-N data. Gaussians fitted to the GCLFs are also plotted.}
 \label{gclfcol}
\end{figure*}

A Gaussian fitted to the GCLF obtained using only GCs at distances
smaller than $10^{\prime}$ ($\sim$ 45 kpc) from the host galaxy is
plotted in Fig. \ref{gauss} (we fit a Gaussian to the binned data). The
fainter part of the GCLF in the shaded region where the completeness is
lower than 50 \% is not used in this fitting process. The turnover
magnitude ($V_{\rm TO}$) and dispersion ($\sigma$) of the GCLF are then
estimated to be $V_{\rm TO} = 23.62 \pm 0.06$ mag and $\sigma = 1.50 \pm
0.04$ mag for the GC population in M87. For the NGC 4552 GCs, $V_{\rm
TO} = 23.56 \pm 0.20$ and $\sigma = 1.34 \pm 0.12$ mag.
In Fig. \ref{gclfcol}, the GCLFs of the red and blue GC subpopulations
are shown with fitted Gaussians. Again the shaded region indicates the
magnitude range where the completeness calculated by using all the GCs
(i.e., blue $+$ red) is lower than 50 \%. This limiting magnitude does
not change a lot if only the red or blue GC subpopulation is considered
since our B- and I-band images are deep enough not to miss a significant
number of red or blue globular clusters detected on the V-band image.
The fitted Gaussians suggest that the GCLF depends on subpopulation;
the $V_{\rm TO}$ of the red GC subpopulation is $\sim 0.5$ mag and
0.4 mag fainter than that of the blue one for M87 and NGC 4552,
respectively, although the results for the NGC 4552 GCs are less
significant due to the large errors.
Larsen et al. (2001) present similar results using the HST/WFPC2 data,
although their $V_{\rm TO}$ values tend to be brighter than those from
other studies. This difference of $V_{\rm TO}$ between the GC
subpopulations is perhaps because of a metallicity difference (Ashman,
Conti \& Zepf 1995; Elson \& Santiago 1996; Jord\'{a}n et al. 2002).
$V_{\rm TO}$ and $\sigma$ of the GCLFs are summarized in Table
\ref{gclfparams}.

\begin{figure*}
 \begin{center}
  \includegraphics[height=14cm,angle=-90,keepaspectratio]{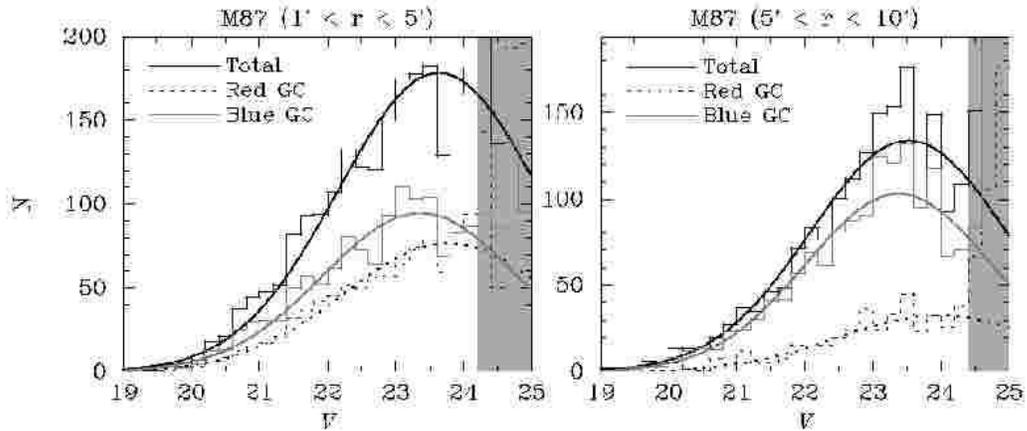}
 \end{center}
 \caption{Same as Fig. \ref{gclfcol}, but the GCLFs in the region of
 $1^{\prime} \leq R \leq 5^{\prime}$ around M87 are compared with those
 in the region of $5^{\prime} \leq R \leq 10^{\prime}$.}
 \label{gclfinout}
\end{figure*}

The GCs within $10^{\prime}$ of M87 are further divided into two samples
at a boundary of $5^{\prime}$ and the GCLFs of all GCs, red GCs, and
blue GCs in the inner and outer regions are presented in Fig.
\ref{gclfinout}. Gaussians are also fitted to these GCLFs and the
$V_{\rm TO}$ and $\sigma$ are summarized in Table \ref{gclfparams}. This
indicates that the shape of GCLF is not significantly different between
the inner region ($1^{\prime} \leq R \leq 5^{\prime}$) and the outer
region ($5^{\prime} \leq R \leq 10^{\prime}$) for red GCs or blue GCs.
Furthermore, the GCLF shape for all GCs is consistent with that in the
core region ($\leq 1^{\prime}$) obtained by Kundu et al. (1999) using
HST/WFPC2: $V_{\rm TO} = 23.67 \pm 0.07$ mag ($\sigma = 1.39 \pm 0.06$).
This suggests that the GCLF shape of the M87 GCs is not a strong
function of distance from the host galaxy. We note that the fainter part
of the GCLF for all GCs tends to be more deficient at the larger
distance; this is due to the lower contribution of the red GC
subpopulation in the outer region. Any radial dependence of the GCLF is
unclear for the GC population around NGC 4552 because the number of GCs
is not substantial enough especially in the outer region ($5^{\prime}
\leq R \leq 10^{\prime}$). Nevertheless, the $V_{\rm TO}$ of the GCLF
obtained at $1^{\prime} \leq R \leq 10^{\prime}$ is consistent with that
in the core region obtained by Kundu \& Whitmore (2001) using HST/WFPC2
who found $V_{\rm TO} = 23.47 \pm 0.14$ mag with a fixed value of
$\sigma = 1.3$, again suggesting that the GCLF shape does not depend
significantly on distance from the host galaxy.

\subsection{GC Specific Frequency}

Using the $V_{\rm TO}$ and $\sigma$ of the Gaussians fitted to the GCLFs
presented in Fig. \ref{gauss}, we can estimate the total number of GCs
associated with M87 or NGC 4552
and calculate a GC specific frequency ($S_N$). We first integrate a GCLF
obtained in an annulus ($1^{\prime}$ width) centered on the host galaxy
down to the magnitude where the completeness becomes 50 \%. This number
is then multiplied by a correction factor to include the fainter GCs
based on the Gaussian GCLF. This calculation is repeated out to a
certain distance from the galaxy centre. We note that it is ideal to fit
a Gaussian to a GCLF obtained in each annulus but this is difficult in
practice, especially at large distances because the number of GCs is not
substantial and statistical errors are significant. We therefore use the
fitted Gaussian in the inner region ($\leq 10^{\prime}$) of M87 or NGC
4552 independently of the distance from the host galaxy. In order to
estimate the number of GCs at distances smaller than $\leq 1^{\prime}$,
where the brightness of host galaxy halo light exceeds the linearity
regime of the CCD, the radial profile of GC surface density outside the
core region is fitted with a de Vaucouleurs law profile (see Paper II
for details) and the fitted formula is extrapolated towards the galaxy
centre. The estimated total number of M87 GCs is $12000 \pm 800$ within
$25^{\prime}$ from the galaxy centre (cf. $13200 \pm 1500$ by HHM). For
NGC 4552, the total number of GCs is estimated to be $1400 \pm 170$
within $10^{\prime}$. Note that the fraction of GCs at $R \leq
1^{\prime}$ estimated from the extrapolation is 8.5 \% for M87 GCs and
17 \% for NGC 4552 GCs. If we adopt $M_V = -22.46$ mag as the $V$-band
absolute magnitude of M87, which is also adopted by HHM (a difference in
adopted distance modulus of 0.03 mag is corrected), the $S_N$ value of
M87 is calculated to be $12.5 \pm 0.8$, which is only slightly smaller
than $14.1 \pm 1.6$ obtained by HHM. The $V$-band luminosity of NGC
4552 is obtained from our data by fitting a de Vaucouleurs law to the
$V$-band surface brightness profile and integrating it out to the same
distance (10$^{\prime}$); $M_V = -21.12$ mag. The $S_N$ value of NGC
4552 is then estimated to be $5.0 \pm 0.6$.\footnote{We calculated the
total number of GCs and $S_N$ within 10$^{\prime}$ for NGC 4552 to avoid
possible contributions of M87 GCs and intergalactic GCs outside of this
radius (see Paper II). If we fit a de Vaucouleurs law to the GC surface
density profile within 10$^{\prime}$ from the NGC 4552 centre and
integrate it out to 25$^{\prime}$ as done for M87 GCs, the number of GCs
and $S_N$ are estimated to be $2000 \pm 660$ and $6.2 \pm 2.1$,
respectively.}

\section{SUMMARY}\label{summary}

We have performed a wide-field imaging survey of the globular cluster
(GC) populations around M87 with Suprime-Cam on the 8.2m Subaru
telescope. A $2^{\circ} \times 0_{\cdot}^{\circ}5$ (560 kpc $\times$ 140
kpc) field extending from M87 to the east was observed through the $BVI$
filters.  In addition to this unprecedented large survey area, our data
analysis has been optimized to study the statistical properties of GCs
as follows:
 
\begin{itemize}
 
 \item GC candidates are isolated not only with an extended source cut
       but also with a colour cut, where only unresolved objects falling
       on a specific region of the $B-V$ and $V-I$ colour-colour diagram
       are accepted. The colour criterion is defined so as to include
       almost all the Galactic GCs and to avoid foreground stars and
       background galaxies. This is expected to efficiently isolate
       bona-fide GCs from other unresolved objects on our imaging data.
 
 \item In order to assess foreground and background contamination which
       needs to be statistically subtracted, we analyze the imaging data
       on the HDF-N field and the Lockman Hole field as control field
       data. These fields cover reasonably wide sky areas ($\sim
       30^{\prime} \times 30^{\prime}$) and are compatible with the data
       of the M87 fields in terms of the filter set ($BVI$), limiting
       magnitudes, and image qualities. We therefore extract
       contaminating populations using identical criteria to those
       adopted in the M87 fields, minimizing the possibility of
       introducing any systematic errors into the subtractive
       correction.
                                                                                
\end{itemize}

In this paper, we have investigated the luminosity function and global
specific frequency ($S_N$) of GC candidates surrounding M87 or NGC 4552.
The $V$-band GC luminosity functions (GCLFs) were obtained in the inner
regions of M87 and NGC 4552 at distances $\leq 10^{\prime}$ from the
galaxy centres. By fitting Gaussians to the GCLFs, the turnover
magnitude is estimated to be $23.62 \pm 0.06$ mag for M87 GCs and $23.56
\pm 0.20$ mag for NGC 4552 GCs. The GCLF appears to depend on GC colour;
the turnover magnitude in the GCLF of the red GC subpopulation ($V-I >
1.1$) is $\sim$ 0.5 mag and 0.4 mag fainter than that of the blue GC
subpopulation ($V-I \leq 1.1$) for the M87 GCs and NGC 4552 GCs,
respectively.

For the M87 GCs, the GCLFs at $1^{\prime} \leq R \leq 5^{\prime}$ were
compared with those at $5^{\prime} \leq R \leq 10^{\prime}$ but no
obvious trend with radius was found in the shape of the GCLF for either
the red or blue subpopulations.
The global $S_N$ of M87 GCs and NGC 4552 GCs is estimated to be 12.5
$\pm$ 0.8 within $25^{\prime}$ and 5.0 $\pm$ 0.6 within $10^{\prime}$,
respectively.

\section*{ACKNOWLEDGEMENTS}

We are grateful to the anonymous referee for careful reading of our
manuscript and helpful comments. This work was based on data collected
at Subaru Telescope and obtained from the SMOKA science archive at
Astronomical Data Analysis Center, which are operated by the National
Astronomical Observatory of Japan. We acknowledge the members of the
Subaru telescope operation team, especially Dr. Hisanori Furusawa for
supports during the observation. This work was partly supported by
Grants-in-Aid for Scientific Research (Nos. 16540223 and 17540216) by
the Japanese Ministry of Education, Culture, Sports, Science and
Technology.

\label{lastpage}


\begin{thebibliography}{}
 
\bibitem[\protect\citeauthoryear{}{}]{}
Ashman, K. M., Conti, A., \& Zepf, S. E. 1995, AJ, 110, 1164

\bibitem[\protect\citeauthoryear{}{}]{}
Ashman, K. M., \& Zepf, S. E. 1992, ApJ, 384, 50

\bibitem[\protect\citeauthoryear{}{}]{}
Baba, H., et al. 2002, ADASS XI, eds. D. A. Bohlender, D. Durand, \&
T. H. Handley, ASP Conference Series, Vol. 281, 298

\bibitem[\protect\citeauthoryear{}{}]{}
Barmby, P. 2003, in Extragalactic Globular Cluster Systems,
	  ed. M. Kissler-Patig (New York, Springer), p. 143.

\bibitem[\protect\citeauthoryear{}{}]{}
Bassino, L. P., Faifer, F. R., Forte, J. C., Dirsch, B., Richtler, T.,
Geisler, D., \& Schuberth, Y. 2006, A\&A, in press (astro-ph/0603349)

\bibitem[\protect\citeauthoryear{}{}]{}
Beasley, M., Baugh, C. M., Forbes, D. A., Sharples, R. M., \& Frenk,
C. S. 2002, MNRAS, 333, 383

\bibitem[\protect\citeauthoryear{}{}]{}
Bertin, E., \& Arnouts, S. 1996, A\&AS, 117, 393

\bibitem[\protect\citeauthoryear{}{}]{}
Blakeslee, J. P. 1999, ApJ, 118, 1506

\bibitem[\protect\citeauthoryear{}{}]{}
Brodie, J. P., Strader, J., Denicol\'{o}, G., Beasley, M. A., Cenarro,
	  A. J., Larsen, S. S., Kuntschner, H., \& Forbes, D. A. 2005,
	  AJ, 129, 2643

\bibitem[\protect\citeauthoryear{}{}]{}
Capak, P., et al. 2004, AJ, 127, 180

\bibitem[\protect\citeauthoryear{}{}]{}
Cohen, J. G., \& Ryzhov, A. 1997, ApJ, 486, 230
                                                                                
\bibitem[\protect\citeauthoryear{}{}]{}
Cohen, J. G., Blakeslee, J. P., \& Ryzhov, A. 1998, ApJ, 496, 808

\bibitem[\protect\citeauthoryear{}{}]{}
C\^{o}t\'{e}, P., Marzke, R. O., \& West, M. J. 1998, ApJ, 501, 554

\bibitem[\protect\citeauthoryear{}{}]{}
de Vaucouleurs, G., \& Nieto, J. -L. 1978, ApJ, 220, 449

\bibitem[\protect\citeauthoryear{}{}]{}
Dirsch, B., Richtler, T., Geisler, D., Forte, J. C., Bassino, L. P., \&
Gieren, W. P. 2003, AJ, 125, 1908

\bibitem[\protect\citeauthoryear{}{}]{}
Elson, R. A. W., \& Santiago, B. X. 1996, MNRAS, 280, 971

\bibitem[\protect\citeauthoryear{}{}]{}
Fern\'{a}ndez-Soto, A., Lanzetta, K. M., \& Yahil, A. 1999, ApJ, 513, 34

\bibitem[\protect\citeauthoryear{}{}]{}
Fioc, M., \& Rocca-Volmerange, B. 1997, A\&A, 326, 950

\bibitem[\protect\citeauthoryear{}{}]{}
Forbes, D. A., Brodie, J. P., \& Grillmair, C. J. 1997, AJ, 113, 1652

\bibitem[\protect\citeauthoryear{}{}]{}
Forbes, D. A., et al. 2004, MNRAS, 355, 608

\bibitem[\protect\citeauthoryear{}{}]{}
Gebhardt, K., \& Kissler-Patig, M. 1999, AJ, 118, 1526

\bibitem[\protect\citeauthoryear{}{}]{}
Geisler, D. 1996, AJ, 111, 480

\bibitem[\protect\citeauthoryear{}{}]{}
Geisler, D., Lee, M. G., \& Kim, E. 1996, AJ, 111, 1529

\bibitem[\protect\citeauthoryear{}{}]{}
Hanes, D. A., C\^{o}t\'{e}, P., Bridges, T. J., McLaughlin, D. E.,
	  Geisler, D., Harris, G. L. H., Hesser, J. E., \& Lee,
	  M. G. 2001, ApJ, 559, 812 (H01)

\bibitem[\protect\citeauthoryear{}{}]{}
Harris, W. E. 1986, AJ, 91, 822

\bibitem[\protect\citeauthoryear{}{}]{}
Harris, W. E. 1991, ARA\&A, 29, 543

\bibitem[\protect\citeauthoryear{}{}]{}
Harris, W. E. 1996, AJ, 112, 1487

\bibitem[\protect\citeauthoryear{}{}]{}
Harris, W. E., Harris, G. L. H., \& McLaughlin, D. E. 1998, AJ, 115,
	  1801 (HHM)

\bibitem[\protect\citeauthoryear{}{}]{}
Jord\'{a}n, A., C\^{o}t\'{e}, P., West, M. J., \& Marzke, R. O. 2002,
	  ApJ {\it Letters}, 576, L113

\bibitem[\protect\citeauthoryear{}{}]{}
Kissler-Patig, M., \& Gebhardt, K. 1998, AJ, 116, 2237

\bibitem[\protect\citeauthoryear{}{}]{}
Kundu, A., Whitmore, B. C., Sparks, W. B., Macchetto, D., Zepf, S. E.,
	  \& Ashman, K. M. 1999, ApJ, 513, 733 (K99)

\bibitem[\protect\citeauthoryear{}{}]{}
Kundu, A., \& Whitmore, B. C. 2001, AJ, 121, 2950

\bibitem[\protect\citeauthoryear{}{}]{}
Landolt, A. U. 1992, AJ, 104, 340

\bibitem[\protect\citeauthoryear{}{}]{}
Larsen, S. S., Brodie, J. P., Huchra, J. P., Forbes, D. A., \&
	  Grillmair, C. J. 2001, AJ, 121, 2974 (L01)

\bibitem[\protect\citeauthoryear{}{}]{}
McLaughlin, D. E. 1999, AJ, 117, 2398

\bibitem[\protect\citeauthoryear{}{}]{}
McLaughlin, D. E., Harris, W. E., \& Hanes, D. A. 1994, ApJ, 422, 486

\bibitem[\protect\citeauthoryear{}{}]{}
Miyazaki, S., et al. 2002, PASJ, 54, 833

\bibitem[\protect\citeauthoryear{}{}]{}
Moore, B., Diemand, J., Madau, P., Zemn M., \& Stadel, J. 2005, MNRAS,
submitted (astro-ph/0510370)

\bibitem[\protect\citeauthoryear{}{}]{}
Ouchi, M., et al. 2004, ApJ, 611, 660

\bibitem[\protect\citeauthoryear{}{}]{}
Peng, E. W., et al. 2005, ApJ, 639, 95

\bibitem[\protect\citeauthoryear{}{}]{}
Pickles, A. 1998, PASP, 110, 863

\bibitem[\protect\citeauthoryear{}{}]{}
Rhode, K. L., \& Zepf, S. E. 2001, AJ, 121, 210

\bibitem[\protect\citeauthoryear{}{}]{}
Rhode, K. L., \& Zepf, S. E. 2004, AJ, 127, 302

\bibitem[\protect\citeauthoryear{}{}]{}
Schlegel, D. J., Finkbeiner, D. P., \& Davis, M. 1998, ApJ, 500, 525

\bibitem[\protect\citeauthoryear{}{}]{}
Strader, J., Brodie, J. P., \& Forbes, D. A. 2004, AJ, 127, 3431

\bibitem[\protect\citeauthoryear{}{}]{}
Strader, J., Brodie, J. P., \& Forbes, D. A. 2004, AJ, 127, 3431

\bibitem[\protect\citeauthoryear{}{}]{}
Strader, J., Brodie, J. P., Spitler, L., \& Beasley, M. A. 2005, AJ,
          submitted (astro-ph/0508001)

\bibitem[\protect\citeauthoryear{}{}]{}
Strom, S. E., Forte, J., Harris, W., Strom, K. M., Wells, D., \& Smith,
	  M. 1981, ApJ, 245, 416

\bibitem[\protect\citeauthoryear{}{}]{}
Tamura, N., Sharples, R. M., Arimoto, N., Onodera, M., Ohta, K., \&
          Yamada, Y. 2006, MNRAS, submitted (Paper II)

\bibitem[\protect\citeauthoryear{}{}]{}
Tonry, J. L., Dressler, A., Blakeslee, J. P., Ajhar, E. A., Fletcher,
	  A. B., Luppino, G. A., Metzger, M. R., \& Moore, C. B. 2001,
	  ApJ, 546, 681

\bibitem[\protect\citeauthoryear{}{}]{}
West, M. J. 1993, MNRAS, 265, 755

\bibitem[\protect\citeauthoryear{}{}]{}
White, R. E. 1987, MNRAS, 227, 185

\bibitem[\protect\citeauthoryear{}{}]{}
Yagi, M., Kashikawa, N., Sekiguchi, M., Doi, M., Yasuda, N., Shimasaku,
K., Okamura, S. 2002, AJ, 123, 66

\end{thebibliography}
\end{document}